\documentclass[acmtodaes]{acmtrans2m}

\usepackage{amsmath}
\usepackage{graphicx}
\usepackage{subfigure}
\usepackage{multirow}
\usepackage{amsfonts}
\usepackage{graphicx}
\usepackage{subfigure}
\usepackage{multirow}

\title{Design of Efficient Reversible Logic Based Binary and BCD Adder Circuits }
\author{ HIMANSHU THAPLIYAL \and
NAGARAJAN RANGANATHAN\\University of South Florida, Tampa}

\begin{abstract}
Reversible logic is gaining significance in the context of emerging technologies such as quantum computing since reversible circuits do not loose information during computation and there is one-to-one mapping between the inputs and outputs. In this work, we present a class of new designs for reversible binary and BCD adder circuits. The proposed designs are primarily optimized for the number of ancilla inputs and the number of garbage outputs and are designed for possible best values for the quantum cost and delay. In reversible circuits, in addition to the primary inputs, some constant input bits are used to realize different logic functions which are referred to as ancilla inputs and are overheads that need to be reduced. Further, the garbage outputs which do not contribute to any useful computations but are needed to maintain reversibility are also overheads that need to be reduced in reversible designs. First, we propose two new designs for the reversible ripple carry adder: (i) one with no input carry $c_0$ and no ancilla input bits, and (ii) one with input carry $c_0$ and no ancilla input bits. The proposed reversible ripple carry adder designs with no ancilla input bits have less quantum cost and logic depth (delay) compared to their existing counterparts in the literature. In these designs, the quantum cost and delay are reduced by deriving designs based on the reversible Peres gate and the TR gate. Next, four new designs for the reversible BCD adder are presented based on the following two approaches: (i) the addition is performed in binary mode and correction is applied to convert to BCD when required through detection and correction, and (ii) the addition is performed in binary mode and the result is always converted using a binary to BCD converter. The proposed reversible binary and BCD adders can be applied in a wide variety of digital signal processing applications and constitute important design components of reversible computing.
\end{abstract}




\begin{document}

\begin{bottomstuff}
Authors' Address: Department of Computer Science and Engineering, 4202 E. Fowler Ave., ENB-118, Tampa, FL-33620, USA; email: \{hthapliy, ranganat\}@cse.usf.edu.
\end{bottomstuff}
\maketitle

\section{Introduction}
In the hardware design, binary computing is preferred over decimal computing because of ease in building hardware based on binary number system \cite{Parhami2}. In spite of ease in building binary hardware, most of the fractional decimal numbers such as 0.110 cannot be exactly represented in binary, thus their approximate values are used for performing computations in binary hardware. Because the financial, commercial, and Internet-based applications cannot tolerate errors generated by conversion between decimal and binary formats, the decimal arithmetic is receiving significant attention and efforts are being accelerated to build dedicated hardware based on decimal arithmetic \cite{Decimal2010}. As the commercial databases tend to contain more decimal than binary data, the use of binary hardware requires the conversion of decimal to binary and vice versa which is an overhead \cite{MFCowlishaw,Akkas}. Recently, software libraries that include conversion capabilities have become available so that the computations appear to be in decimal making it transparent to the programmer. But the software implementation of decimal arithmetic is usually 100 to 1000 times slower than implementing in hardware \cite{Decimal20101}. 

Among the emerging computing paradigms, reversible logic appears to be promising due to its wide applications in emerging technologies such as quantum computing, quantum dot cellular automata, optical computing, etc \cite{Nielsen,XMa,Opticalreversible,Parhami1,XMa1}. Reversible logic is also being investigated for its promising applications in power-efficient nanocomputing\cite{IEEE2010,Frank1}. Reversible circuits are those circuits that do not lose information during computation and reversible computation in a system can be performed only when the system comprises of reversible gates. These circuits can generate unique output vector from each input vector, and vice versa, that is, there is a one-to-one mapping between the input and the output vectors.  

 A quantum computer will be viewed as a quantum network (or a family of quantum networks) composed of quantum logic gates; each gate performing an elementary unitary operation on one, two or more two-state quantum systems called qubits. Each qubit represents an elementary unit of information; corresponding to the classical bit values 0 and 1. Any unitary operation is reversible and hence quantum networks must be built from reversible logical components\cite{Nielsen,Vedral}. \emph{ Quantum computers  of many qubits are extremely difficult to realize thus the number of qubits  in the quantum circuits needs to be minimized \cite{Takahashi_survey,Takahashi2005}. This sets the major objective of optimizing the number of ancilla input qubits and the number of the garbage outputs in the reversible logic based quantum circuits}. The constant input in the reversible quantum circuit is called the ancilla input qubit (ancilla input bit), while the garbage output refers to the output which exists in the circuit just to maintain one-to-one mapping but is neither one of the primary inputs nor a useful output. Thus, the inputs regenerated at the outputs are not considered as garbage outputs \cite{Fredkin}.

The proposed work focuses on the design of  reversible binary and the
BCD adder circuits primarily optimized for number of ancilla input
bits and the garbage outputs. As the optimization of ancilla
input bits and the garbage outputs may impact the design in
terms of the quantum cost and the delay, thus quantum cost
and the delay parameters are also considered for optimization
with primary focus towards the optimization of number of
ancilla input bits and the garbage outputs.
To the best of our knowledge this is the first attempt in the literature that explores the reversible BCD adder designs  with the goal of optimizing the number of ancilla input bits and the garbage outputs. First, we propose two new designs for the reversible ripple carry adder: (i) one with no input carry $c_0$ and no ancilla input bits, and (ii) one with input carry $c_0$ and no ancilla input bits. The proposed reversible ripple carry adder designs with no ancilla input bits have less quantum cost and logic depth (delay) compared to their existing counterparts in the literature. In these designs, the quantum cost and delay are reduced by deriving designs based on the reversible Peres gate and the TR gate. Next, four new designs for the reversible BCD adder are presented based on the following two approaches: (i) the addition is performed in binary mode and correction is applied to convert to BCD when required through detection and correction, and (ii) the addition is performed in binary mode and the result is always converted using a binary to BCD converter.  The various reversible components needed in the BCD adder design  are optimized in parameters of number of ancilla input bits/qubits and the number of garbage outputs and explore the possible best values for the quantum cost and delay.  The comparison of the proposed designs with the existing designs is also illustrated. 

The paper is organized as follows: The background of the reversible logic and the details of the existing works are presented in Section II; the improved design of the TR gate is illustrated in Section III.  The design methodologies of proposed reversible ripple carry adder with no input carry and with input carry are discussed in Section IV and V, respectively. The designs of the reversible BCD adder are addressed in Section VI. The simulation and the verification of the proposed designs using Verilog HDL are presented in Section VII while the conclusions are provided in Section VIII.

\section{Background}
The most popular reversible gates are the Fredkin gate \cite{Fredkin}, the Toffoli gate \cite{Toffoli1} and the Peres gate \cite{Peres}. Each reversible gate has an associated implementation cost called the quantum cost \cite{Smolin}. The quantum cost of a reversible gate is the number of 1x1 and 2x2 reversible gates or quantum logic gates required in its design. The quantum costs of all reversible 1x1 and 2x2 gates are taken as unity \cite{Smolin,HungTCAD06,Maslov1}. The 3x3 reversible gates are  realized  using 1x1 NOT gate, and 2x2 reversible gates such as Controlled-V and Controlled-$V^{+}$ (V is a square-root-of NOT gate and $V^{+}$ is its hermitian) and the Feynman gate  which is also known as the Controlled NOT gate (CNOT). The quantum cost of a reversible gate can be calculated by counting the numbers of NOT, Controlled-V, Controlled-$V^{+}$ and CNOT gates required in its implementation.

\subsection {The NOT Gate}
A NOT gate is a 1x1 gate represented as shown in Fig. \ref{NOTgate}. Since it is a 1x1 gate, its quantum cost is unity.

\subsection {The Controlled-V and Controlled-$V^{+}$ Gates}
\begin{figure}[!h]
 \begin{center}
\subfigure[NOT Gate]
   {\label{NOTgate}
   \includegraphics[width=1.5in]{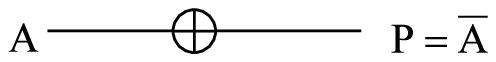}}
   \quad
 \subfigure[Controlled-V Gate]
   {\label{VGate}
   \includegraphics[width=1.5in]{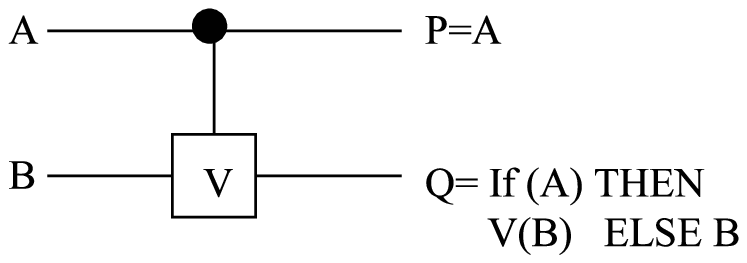}}
   \quad
    \subfigure[Controlled-$V^{+}$ Gate ]
   {\label{V+gate}
   \includegraphics[width=1.5in]{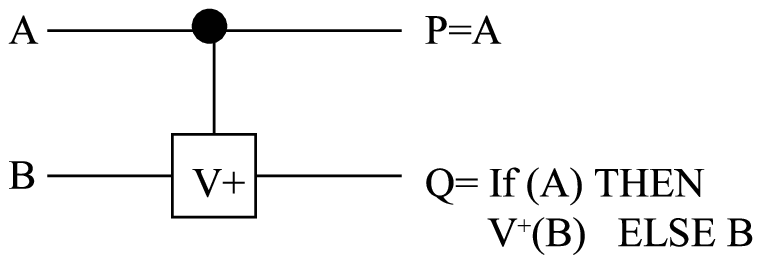}}
   \quad
  
 \end{center}
 \caption{ The Controlled-V and Controlled-$V^{+}$ Gates}
 \label{ControlledVV+gate}
\end{figure}

The controlled-V gate is shown in Fig. \ref{VGate}. In the controlled-V gate, when the control signal A=0 then the qubit B will pass through the controlled part unchanged, i.e., we will have Q=B. When A=1  then the unitary operation $V = \frac{i+1}{2} \left( \begin{smallmatrix} 1&-i\\ -i&1 \end{smallmatrix} \right)$ is applied to the input B, i.e., Q=V(B). The controlled-$V^{+}$ gate is shown in Fig. \ref{V+gate}. In the controlled-$V^{+}$ gate when the control signal A=0 then the  qubit B will pass through the controlled part unchanged, i.e., we will have Q=B. When A=1  then the unitary operation $V^{+}$ = $V^{-1}$  is applied to the input B, i.e., Q=$V^{+}$(B).
   
The V and $V^{+}$ quantum gates have the following properties: 
\[
\begin{array}{rclcrcl}
V \times V= NOT \\
V \times V^{+} =  V^{+} \times V  =I \\
 V^{+} \times V^{+} = NOT
\end{array}
\]

The properties above show that when two V gates are in series they will behave as a NOT gate. Similarly, two $V^{+}$ gates in series also function as a NOT gate.  A V gate in series with $V^{+}$ gate, and vice versa, is an identity. For more details of the V and $V^{+}$ gates, the reader is referred to \cite{Nielsen,HungTCAD06,Maslov1}.

\begin{figure}[!h]
 \begin{center}
 \subfigure[ CNOT Gate ]
   {\label{Feynmangate}
   \includegraphics[width=1.5in]{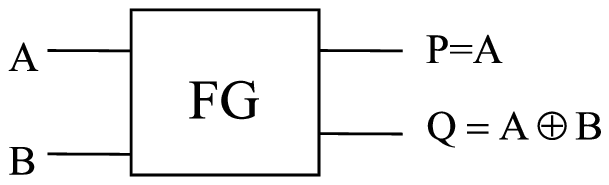}}
   \quad
    \subfigure[Quantum representation ]
   {\label{CNTgate}
   \includegraphics[width=1.5in]{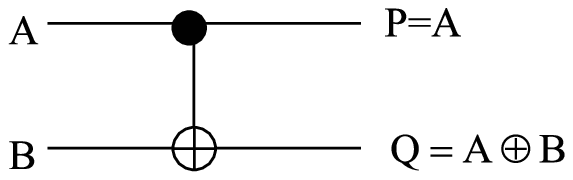}}
   \quad
  
 \end{center}
 \caption{ The CNOT gate and its quantum representation}
 \label{NCNTgate}
\end{figure}

\subsection{The Feynman Gate (CNOT Gate)}
The Feynman gate (FG) or the Controlled-NOT gate (CNOT) is a 2 inputs 2 outputs reversible gate having the mapping (A, B) to (P=A, Q=A $\oplus$ B) where A, B are the inputs and P, Q are the outputs, respectively. Since it is a 2x2 gate, it has a quantum cost of 1. Figures \ref{Feynmangate} and \ref{CNTgate} show the block diagrams and quantum representation of the Feynman gate.
  
\subsection {The Toffoli Gate}
\begin{figure}[!h]
 \begin{center}
 \subfigure[Toffoli Gate]
   {\label{ToffoliGate}
   \includegraphics[width=1.8in]{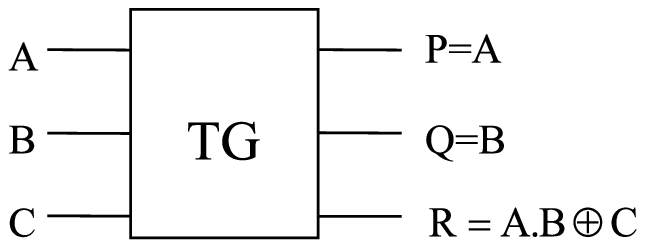}}
   \quad
\subfigure[Quantum implementation of Toffoli gate]
   {\label{ToffliGate_delay}
   \includegraphics[width=2.0in]{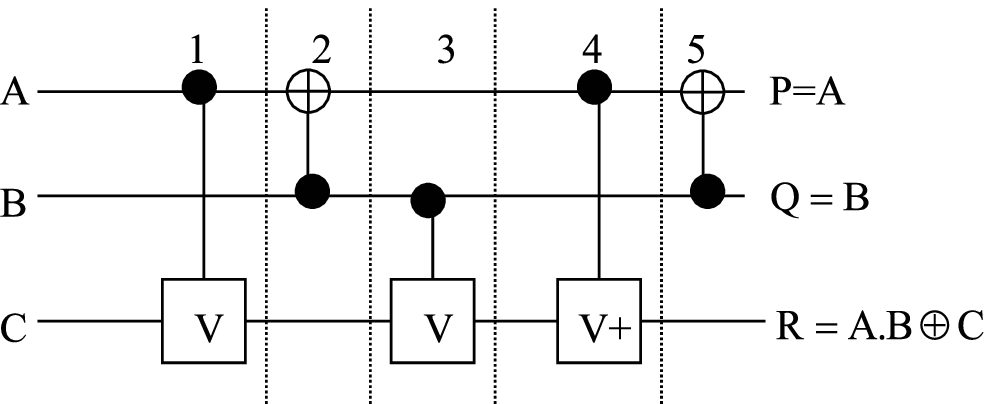}}
   \quad
      
 \end{center}
 \caption{ The Toffoli gate and its quantum implementation}
 \label{Toffoli_Gate}
\end{figure}

The Toffoli Gate (TG) is a 3x3 two-through reversible gate as shown in Fig. \ref{ToffoliGate}. Two-through means two of its outputs are the same as the inputs with the mapping (A, B, C) to (P=A, Q=B, R=A$\cdot$B$\oplus$ C), where A, B, C are inputs and P, Q, R are outputs, respectively. The Toffoli gate is one of the most popular reversible gates and has the quantum cost of 5 as shown in Fig. \ref{ToffliGate_delay} \cite{Toffoli1}. The quantum cost of Toffoli gate is 5 as it needs 2V gates, 1 $V^{+}$ gate and 2 CNOT gates to implement it.

\subsection{The Peres Gate}
\begin{figure}[!h]
 \begin{center}
 \subfigure[Peres Gate]
   {\label{PeresGate}
   \includegraphics[width=1.8in]{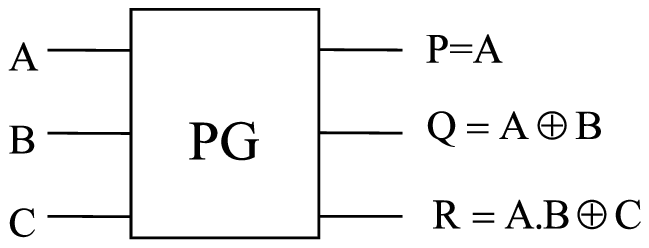}}
   \quad
\subfigure[Quantum implementation of Peres gate]
   {\label{PeresGate_delay}
   \includegraphics[width=2.0in]{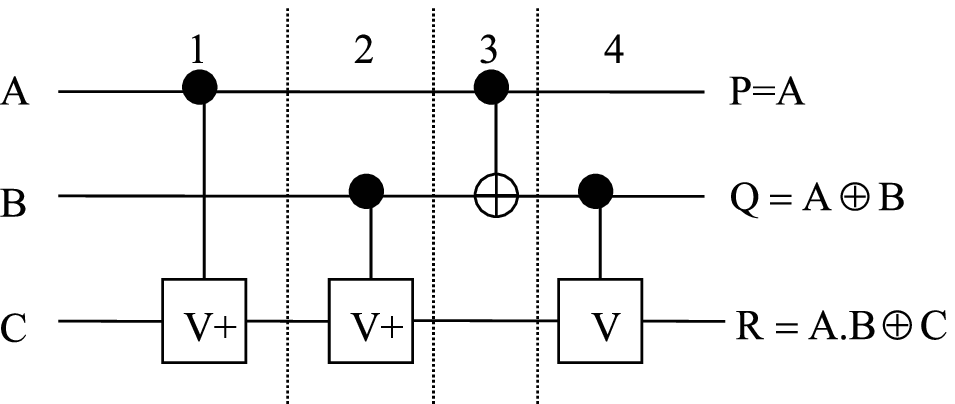}}
   \quad
      
 \end{center}
 \caption{ The Peres gate and its quantum implementation}
 \label{Peres_Gate}
\end{figure}

The Peres gate is a 3 inputs 3 outputs (3x3) reversible gate having the mapping (A, B, C) to (P=A, Q=A$\oplus$B, R= A$\cdot$B$\oplus$C), where A, B, C are the inputs and P, Q, R are the outputs, respectively \cite{Peres}. Figure \ref{PeresGate} shows the Peres gate and Fig. \ref{PeresGate_delay} shows the quantum implementation of the Peres gate (PG) with quantum cost of 4 \cite{HungTCAD06}. The quantum cost of Peres gate is 4 since it requires 2 $V^{+}$ gates, 1 V gate and 1 CNOT gate in its design.  In the existing literature, among the 3x3 reversible gates, the Peres gate has the minimum quantum cost. 

\subsection {Delays of the Reversible Gates}
Delay is another important parameter that can indicate the efficiency of the reversible circuits. Here, delay represents the critical delay of the circuit. In our delay calculations, we use the logical depth as the measure of the delay \cite{Mohammadi_merit}. The delays of all 1x1 gate and 2x2 reversible gate are taken as unit delay called $\Delta$. Any 3x3 reversible gate can be designed  from 1x1 reversible gates and  2x2 reversible gates, such as the CNOT gate, the Controlled-V and the Controlled-$V^{+}$ gates. Thus the delay of a 3x3 reversible gate can be computed by calculating its logical depth when it is designed from smaller 1x1 and 2x2 reversible gates. Figure \ref{ToffliGate_delay} shows the logic depth in the quantum implementation of Toffoli gate. Thus, it can be seen that the Toffoli gate has the delay of 5 $\Delta$. Each 2x2 reversible gate in the logic depth contributes to 1 $\Delta$ delay. Similarly, Peres gate shown in Fig. \ref{PeresGate_delay} has the logic depth of 4 that results in its delay as 4 $\Delta$. 

\subsection{Prior Works}
The research on reversible logic is expanding towards both design and synthesis. In the synthesis of reversible logic circuits there has been several interesting attempts in the literature such as in \cite{Jha1,Shende03synthesisof,MaslovTCAD04,PerkowskiCJ08,Markov}. The researchers have addressed the optimization of reversible logic circuits from the perspective of quantum cost and the number of garbage outputs. Recently, in \cite{Wille1,Wille2} interesting contributions are made toward deriving exact minimal elementary quantum gate realization of reversible combinational circuits.  Thus, in synthesis of reversible logic circuits the optimization in terms of number of ancilla input bits and also the delay are not yet addressed except the recent work in \cite{WilleDAC10} which discusses about the post synthesis method for reducing the number of lines (qubits) in the reversible circuits. The designs of reversible sequential circuits are also addressed in literature in which various latches, flip-flops, etc. are designed \cite{Rice,Chuang,Sastry,ThapliyalJETC,Thapliyal3}. 

Reversible arithmetic units such as adders, subtractors, multipliers which form the essential component of a computing system have also been designed in binary as well as ternary logic such as in \cite{vos_adder,Haghparastmul,Biswas,Bruce02efficientadder,Khan,Khan2}. In \cite{Draper_ripple}, researchers have designed the quantum ripple carry adder having no input carry with one ancilla input bit. In \cite{Takahashi2005,Takahashi_ripple}, the researchers have investigated  new designs of the quantum ripple carry adder with no ancilla input bit and improved delay. In \cite{Meter1}, the measurement based design of carry look-ahead adder is presented while in \cite{Meter2} the concept of arithmetic on a distributed-memory quantum multicomputer is introduced.  A comprehensive survey of quantum arithmetic circuits can be found in \cite{Takahashi_survey}. 

The design of BCD adders and subtractors have also been attempted. The researchers have investigated the design of BCD adders and subtractors in which parameters such as the number of reversible gates, number of garbage outputs, quantum cost, number of transistors, etc  are considered for optimization \cite{babu_bcd06,Biswas,Majid_bcd1,Thomsen,majid_bcd2009,jamesbcd}. Thus to the best of our knowledge researchers have not yet addressed the design of the BCD arithmetic units primarily focusing on optimizing the number of ancilla input bits and the garbage outputs. In this work, we present a class of new designs for reversible binary and \emph{BCD adder circuits}. The proposed designs are primarily optimized for the number of ancilla inputs and the number of garbage outputs and are designed for  possible best values for the quantum cost and delay.

\section{Proposed Design of the TR gate}
\begin{figure}[!h]
 \begin{center}
   \subfigure[Quantum symbol of the TR Gate]
   {\label{TR_Gate}
   \includegraphics[width=1.7in]{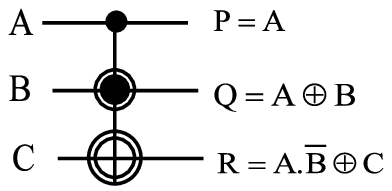}}
   \quad
\subfigure[Quantum realization of the TR gate]
   {\label{TRGate_quant}
   \includegraphics[width=2.25in]{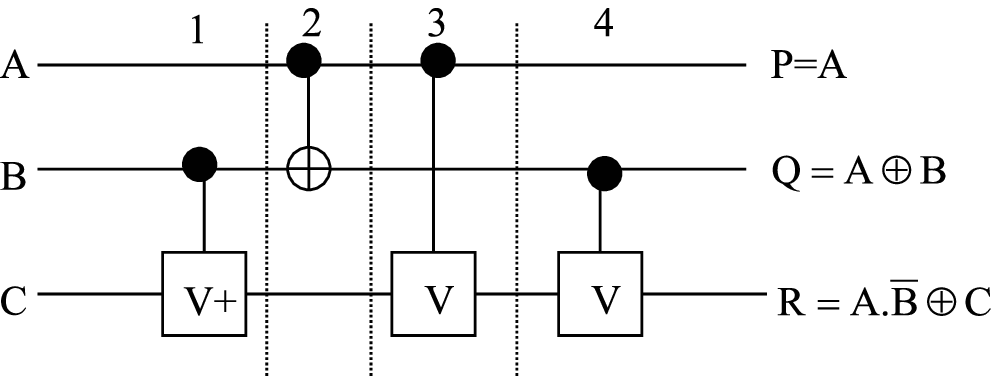}}
   \quad
 \end{center}
 \caption{ TR gate and its improved quantum implementation}
 \label{TR+gate}
\end{figure}

The reversible TR gate is a 3 inputs 3 outputs gate having inputs to outputs mapping as (P=A, Q=A $\oplus$ B, $R=A \cdot \bar{B} \oplus C$ \cite{ThapliyalISVLSI}. We present the graphical notation of the TR gate in Fig. \ref{TR_Gate} along with its new quantum implementation with 2x2 quantum gates  in Fig. \ref{TRGate_quant}.  The TR gate is designed from 1 Controlled $V$ gate, 1 CNOT gate, and 2 Controlled $V^{+}$ gates resulting in its quantum cost as 4. Further, the logic depth of the quantum implementation of the TR gate is 4 resulting in its propagation delay as 4 $\Delta$. The quantum cost and the delay of the TR gate was earlier estimated as 6 and 6 $\Delta$, respectively \cite{ThapliyalISVLSI}. The TR gate can realize the Boolean functions $A \cdot \bar{B} \oplus C$ and A $\oplus$ B with only gate. Further, it can implement the functions such as $A \cdot \bar{B}$ when its input C is tied to 0. These properties of the TR gate make it very useful in designing the reversible arithmetic units \cite{ThapliyalDATE,ThapliyalNANO1}.

\section{Design Methodology of Proposed Reversible ripple carry adder with no input carry}
We present the design of reversible ripple carry adder with no input carry($c_0$) and is designed without any ancilla inputs and the garbage outputs. \emph{The proposed method  improves the quantum cost and the delay of the reversible ripple carry adder compared to the existing design approaches which have optimized the adder design in terms of number of ancilla inputs}. Consider the addition of two n bit numbers  $a_{i}$ and $b_{i}$ stored at memory locations $A_{i}$ and $B_{i}$, respectively, where  $0\le i \le n-1$. Further, consider that memory location  $A_{n}$ is initialized with z $\in$ \{0, 1\}.  At the end of the computation, the memory location $B_{i}$ will have  $s_{i}$, while the location $A_{i}$ keeps the value $a_{i}$. The additional location $A_{n}$ that initially stores the value z  will  have the value  $z\oplus s_{n}$  at the end of the computation. Thus $A_{n}$  will have the value of $s_{n}$ when z=0.   Here, $s_{i}$ is the sum bit produced and is defined as: \\
$
     s_{i} = 
\begin{cases} 
 a_{i}\oplus b_{i} \oplus c_{i} & \mbox{if  $0\le i\le n-1$} \\
c_{n} & \mbox{if $i=n$} 
\end{cases}
$\\
where $c_{i}$ is the carry bit and is defined as:\\
$
     c_{i} = 
\begin{cases} 
 0 & \mbox{if } i=0 \\
 a_{i-1} b_{i-1} \oplus b_{i-1} c_{i-1} \oplus c_{i-1} a_{i-1} & \mbox{if } 1\le i\le n
\end{cases}
$

The proposed design methodology of generating the reversible ripple carry adder with no input carry minimizes the garbage  outputs by producing the carry bits $c_{i}$ based on the inputs  $a_{i-1}$, $b_{i-1}$ and the carry bit $c_{i-1}$ from the previous stage. Once all the carry bits $c_{i}$ are generated they are stored at memory location $A_{i-1}$ which was initially used for storing the input $a_{i-1}$ for $0\le i\le n-1$. After the generated carry bits are used for further computation, the location $A_{i}$ are restored to the value $a_{i}$ while the location $B_{i}$ stores the sum bit $s_{i}$ for  $0 \le i \le n-1$. Thus restoring of location $A_{i}$ to the value $a_{i}$ helps in minimizing the garbage outputs. Since no constant input having the value as 0 is needed in the proposed approach, it saves the ancilla inputs.  The proposed methodology of generating the reversible ripple adder circuit without input carry is referred as methodology 1 in this work. The proposed methodology is generic in nature and can design the reversible ripple carry adder circuit with no input carry of any size.  The steps involved in the proposed methodology is explained for addition of two n bit numbers  $a_{i}$ and $b_{i}$, where  $0\le i \le n-1$.  An illustrative example of generation of reversible ripple carry adder circuit that can perform the addition of two 8 bit  numbers $a$=$a_0...a_7$ and $b$=$b_0...b_7$ is also shown.\\\\

\textsf{Steps of Methodology 1 (Reversible Adder Circuit With No Input Carry)}

\begin{figure*}[!ht]
 \begin{center}
 \subfigure[After Step 1]
   {\label{8bitadder0ancilla1}
   \includegraphics[width=0.8in]{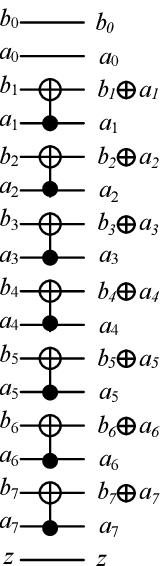}}
   \quad
\subfigure[After Step 2]
   {\label{8bitadder0ancilla2}
   \includegraphics[width=2.25in]{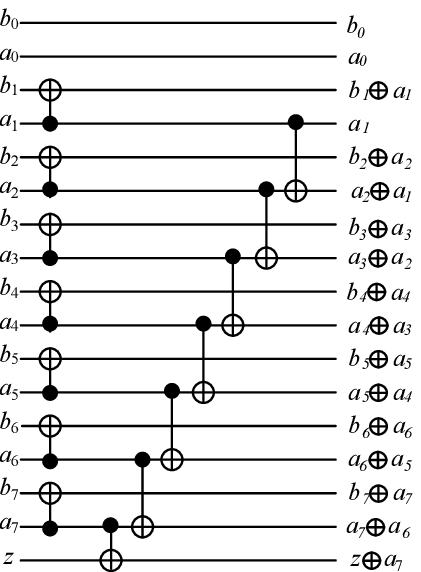}}
   \quad
\subfigure[After Step 3]
   {\label{8bitadder0ancilla3}
   \includegraphics[width=2.75in]{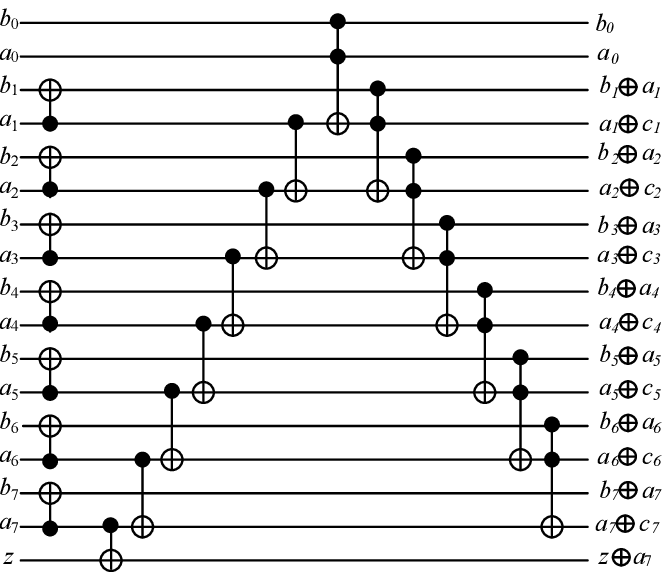}}
   \quad
    
 \end{center}
 \caption{Circuit generation of reversible 8 bit adder with no input carry: Steps 1-3}
 \label{84bitadder0ancilla1-3}
\end{figure*}

\begin{figure*}[!ht]
 \begin{center}
 \subfigure[After Step 4]
   {\label{8bitadder0ancilla4}
   \includegraphics[width=3.5in]{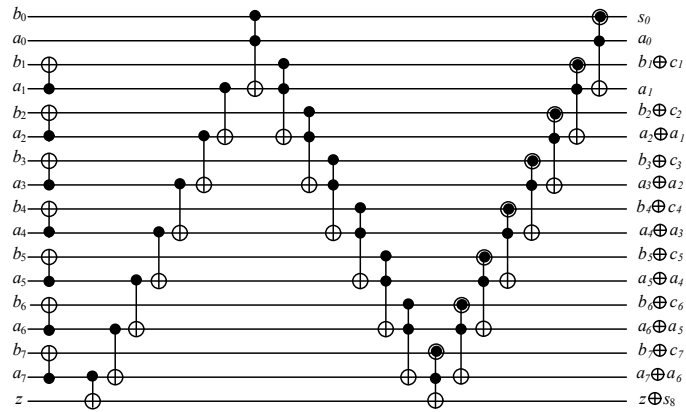}}
   \quad
\subfigure[After Step 5]
   {\label{8bitadder0ancilla5}
   \includegraphics[width=3.5in]{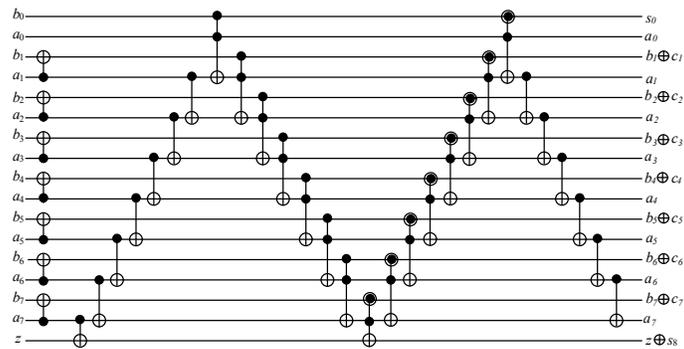}}
   \quad
\subfigure[After Step 6]
   {\label{8bitadder0ancilla}
   \includegraphics[width=3.5in]{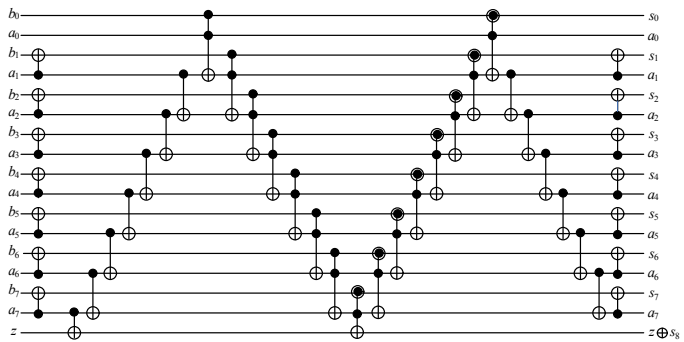}}
   \quad
      
 \end{center}
 \caption{ Circuit generation of reversible 8 bit adder with no input carry: Steps 4-6}
 \label{84bitadder0ancilla4-6}
\end{figure*}

\begin{enumerate}
\item For i=1 to n-1:\\ 
At pair of locations $A_i$ and $B_i$ apply the CNOT gate such that the location $A_i$ will maintain the same value, while location $B_i$ transforms to (*$A_{i}$ $\oplus$ *$B_{i}$), where *$A_{i}$ and *$B_{i}$ represent the values stored at location $A_{i}$ and $B_{i}$.  The step 1 is shown for reversible ripple carry adder circuit that can perform the  addition of two 8 bit numbers in Fig.\ref{8bitadder0ancilla1}.  

\item  For i=n-1 to 1: \\
At pair of locations $A_i$ and $A_{i+1}$ apply the CNOT gate such that the location $A_i$ will maintain the same value, while the location  $A_{i+1}$ transforms to  (*$A_{i}$ $\oplus$ *$A_{i+1}$). The step 2 is shown for reversible 8 bit adder circuit in Fig.\ref{8bitadder0ancilla2}.

\item For i=0 to n-2: \\
At locations $B_i$, $A_i$ and and $A_{i+1}$ apply the Toffoli gate such that $B_i$, $A_i$ and and $A_{i+1}$ are passed to the inputs A, B, C, respectively, of the Toffoli gate. The step 3 is shown for reversible 8 bit adder circuit in Fig.\ref{8bitadder0ancilla3}.  

\item For i=n-1 to 0: \\
At locations $A_i$, $B_i$ and and $A_{i+1}$ apply the Peres gate such that $A_i$, $B_i$ and and $A_{i+1}$ are passed to the inputs A, B, C, respectively, of the Peres gate. The step 4 is shown for reversible 8 bit adder circuit in Fig.\ref{8bitadder0ancilla4}. 

\item  For i=1 to n-2: \\
At pair of locations $A_i$ and $A_{i+1}$ apply the CNOT gate such that the location $A_i$ will maintain the same value, while location $B_i$ transforms to the value  (*$A_{i}$ $\oplus$ *$B_{i}$).  The step 5 is shown for reversible 8 bit adder circuit in Fig.\ref{8bitadder0ancilla5}. 

\item For i=1 to n-1 : \\
At pair of locations $B_i$ and $A_{i}$ apply the CNOT gate such that the location $A_i$ will maintain the same value, while location $B_i$ transforms to the value  (*$A_{i}$ $\oplus$ *$b_{i}$).  This final step will result in a reversible adder circuit that can perform the  addition of two n bit numbers. For reversible 8 bit adder circuit, the design is shown in Fig.\ref{8bitadder0ancilla}. 

\end{enumerate}

\begin{figure}[!ht]
\begin{center}
\includegraphics[width=3.0in]{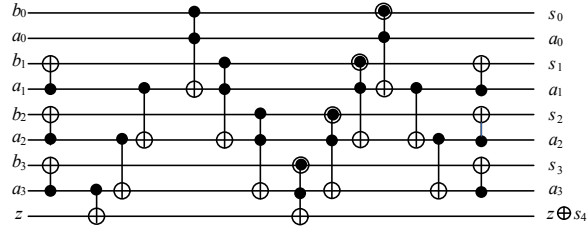}
\end{center} 
\caption{Proposed reversible 4 bit adder without input carry}
\label{4bitadder0ancilla}
\end{figure}

\emph{Thus, the proposed methodology implements the reversible ripple carry adder with no input carry, without any ancilla input bit}. Since in the design of the reversible BCD adder, the 4 bit reversible ripple carry adder will be used, thus its design is also illustrated in Fig.\ref{4bitadder0ancilla}\\

\textsf{Theorem 1}: \emph{Let a and b are two n bit binary numbers represented as  $a_{i}$ and $b_{i}$ and z $\in$ \{0, 1\} is another 1 bit input, where $0 \le i \le n-1$, then the proposed design steps of methodology 1 result in the ripple carry adder circuit that works correctly. The proposed design methodology  designs an n bit adder circuit that  produces the sum output $s_{i}$ at the memory location where $b_{i}$ is stored, while restores the location where  $a_{i}$ is initially~stored  to the value $a_{i}$ for  $0 \le i \le n-1$. Further, the proposed design methodology transforms the memory location where z is initially  stored  to  $z\oplus s_{n}$, and restores the memory  location where the input carry $c_0$ is initially stored to the value $c_0$}.\\\\

\textsf{Proof}:  The proposed approach will make the following changes on the inputs that are illustrated as follows:
\begin{enumerate}
\item Step 1: The step 1 of the proposed approach transforms the input states to  \\
$\left | b_0 \right \rangle \left | a_0 \right \rangle \left (\bigotimes _{i=1}^{n-1} \left | b_i \oplus a_i \right \rangle  \left | a_i \right \rangle   \right )  \left | z  \right \rangle$ \\

An example of the transformation of the input states after step 1 is illustrated for 8 bit reversible ripple carry adder circuit in Fig.\ref{8bitadder0ancilla1}.  

\item Step 2:  The step 2 of the proposed approach transforms the input states to \\
\begin{eqnarray*}
\left | b_0 \right \rangle \left | a_0 \right \rangle \left | b_1\oplus a_1 \right \rangle \left | a_1 \right \rangle\left (\bigotimes _{i=2}^{n-1} \left | b_i \oplus a_i \right \rangle  \left | a_i \oplus a_{i-1}\right \rangle   \right ) \left   | z \oplus a_{n-1} \right \rangle
\end{eqnarray*}

An example of the  transformation of the input states after step 2 is illustrated for 8 bit reversible ripple carry adder circuit in Fig.\ref{8bitadder0ancilla2}.  

\item Step 3: 
The step 3 has n-1 Toffoli gates. The first Toffoli gate takes the inputs as $b_0$,$a_0$ and $a_1$ and produces the output as $b_0$,$a_0$ and $a_1\oplus c_1$. The third output of the Toffoli gate produces $a_1\oplus c_1$ because $c_1$=$a_0 \cdot b_0$ where $c_1$ represents the generated output carry after addition of $a_0$ and $b_0$. 
The remaining n-2 Toffoli gates take the inputs as  $b_i\oplus a_i$, $a_i\oplus c_i$, $a_i\oplus a_{i+1}$ and produces the outputs as  $b_i\oplus a_i$, $a_i\oplus c_i$, $a_{i+1}\oplus c_{i+1}$ where $1 \le i \le n-1$. Thus, after the step 3, the input states is transformed to \\
 
$\left | b_0 \right \rangle \left | a_0 \right \rangle \left (\bigotimes _{i=1}^{n-1} \left | b_i \oplus a_i \right \rangle  \left | a_i \oplus c_i \right \rangle   \right )  \left | z \oplus a_{n-1} \right \rangle $ \\

An example of the transformation of the input states after step 3 is illustrated for 8 bit reversible ripple carry adder circuit in Fig.\ref{8bitadder0ancilla3}.  

\item Step 4:  The Step 4 has n Peres gates. The n-1 Peres gate take the inputs as $a_i \oplus c_{i}$, $b_i \oplus a_{i}$, $a_{i+1} \oplus c_{i+1}$ to produce the outputs as  $a_i \oplus c_{i}$, $b_i \oplus c_{i}$, $a_{i} \oplus a_{i+1}$. The third outputs of the Peres gate are $a_{i} \oplus a_{i+1}$ because it realizes the function $A\cdot B \oplus C$ where A, B and C are the inputs of the Peres gate. Hence the Peres gates will have the third outputs as  $ a_i \oplus c_{i} \cdot b_i \oplus c_{i} \oplus a_{i} \oplus a_{i+1}$= $a_{i} \oplus a_{i+1}$.  The nth Peres gate takes the inputs as  $a_0$, $b_0$, $a_{1} \oplus c_{1}$ to produce the outputs as  $a_0$, $a_0 \oplus b_0$, $a_{1}$. Please note that $s_0=a_0 \oplus b_0$. Thus the step 4 transforms the input states to\\
\begin{eqnarray*}
\left | s_0 \right \rangle \left | a_0 \right \rangle \left | b_1\oplus c_{1} \right \rangle \left | a_1 \right \rangle\left (\bigotimes _{i=2}^{n-1} \left | b_i \oplus c_i \right \rangle  \left | a_i \oplus a_{i-1}\right \rangle   \right ) \left | z \oplus s_{n} \right \rangle
\end{eqnarray*} \\

An example of the transformation of the input states after step 4 is illustrated for 8 bit reversible ripple carry adder circuit in Fig.\ref{8bitadder0ancilla4}.  

\item Step 5: The step 5 of the proposed approach transforms the input states to \\

$\left | s_0 \right \rangle \left | a_0 \right \rangle \left (\bigotimes _{i=1}^{n-1} \left | b_i \oplus c_i \right\rangle  \left | a_i \right \rangle   \right )  \left | z \oplus s_{n} \right \rangle$\\

An example of the transformation of the input states after step 5 is illustrated for 8 bit reversible ripple carry adder circuit in Fig.\ref{8bitadder0ancilla5}.  

\item Step 6: The step 6 of the proposed approach transforms the input states to
\\
$\left (\bigotimes _{i=0}^{n-1} \left | s_i \right \rangle  \left | a_i \right \rangle   \right )  \left | z \oplus s_{n} \right \rangle$ \\

An example of the transformation of the input states after step 6 is illustrated for 8 bit reversible ripple carry adder circuit in Fig.\ref{8bitadder0ancilla}.  

\end{enumerate} 

Thus, the proposed six steps transform the memory location  where $b_{i}$ is initially stored  to the sum output $s_{i}$, while the location where  $a_{i}$ is initially stored will be restored to the value $a_{i}$ for  $0 \le i \le n-1$ after the generation of the output carries and their subsequent use to produce the sum outputs. The memory location where z is stored will have  $z\oplus s_{n}$ and  the memory  location where the input carry $c_0$ was stored initially will be restored to the value $c_0$.  In summary, the proposed design methodology 1 generates the n bit reversible ripple carry adder that is functionally correct.   

\emph{ Delay and Quantum Cost}
\begin{itemize}
\item Step 1 of the proposed methodology needs $n-1$ CNOT gates working in parallel thus this step has the quantum cost of $n-1$ and delay of $1 \Delta$.
\item  Step 2 of the proposed methodology needs $n$ CNOT gates working in series thus this step has the quantum cost of $n$ and delay of $n \Delta$.
\item  Step 3 needs $n-1$ Toffoli gates working in series thus this step has the quantum cost of $5(n-1)$ and delay of $5(n-1) \Delta$.
\item  Step 4 needs $n$ Peres gates working in series thus this step has the quantum cost of $4n$ and delay of $4n \Delta$.
\item  Step 5 needs $n-1$ CNOT gates working in series thus this step has the quantum cost of $n-1$ and delay of $(n-1) \Delta$.
\item  Step 6 needs $n-1$ CNOT gates working in parallel thus this step has the quantum cost of $n-1$ and delay of $1 \Delta$.
\end{itemize}
Thus, the total quantum cost of n bit reversible ripple carry adder is $n-1+n+5(n-1)+4n+n-1+n-1 = 13n-8$. The propagation delay will be $1 \Delta+n \Delta+5(n-1) \Delta+4n \Delta+(n-1) \Delta+1 \Delta = (11n-4) \Delta$.
A comparison of the proposed design with the existing designs is illustrated in Table \ref{tabSR2}. In Table \ref{tabSR2}, for \cite{Takahashi2005} the quantum cost and the delay values are valid for $n\ge 2$, and for n=1 the design has the quantum cost of 8 and delay of 8 $\Delta$.    Among the existing designs of the reversible ripple carry adder with no input carry, the designs in \cite{Takahashi2005} and \cite{Takahashi_ripple} are designed with no ancilla input bits and the garbage outputs, while the design presented in \cite{Draper_ripple} has 1 ancilla input and 1 garbage output. In this work we have compared our proposed design of the reversible carry adder with the designs in \cite{Draper_ripple}, \cite{Takahashi2005} and \cite{Takahashi_ripple} for values of n varying from 8 bits to 512 bits. Table \ref{tabSR2QC} shows the comparison in terms of quantum cost which shows that the proposed design of the reversible carry adder with no input carry achieves the  improvement ratios ranging from 22.5\% to 23.51\%, 46.36\% to 49.95\%, and 13.37\% to 13.33\% compared to the design presented in \cite{Draper_ripple}, \cite{Takahashi2005} and \cite{Takahashi_ripple}, respectively. From Table \ref{tabSR2D}, it can be seen that the proposed design of reversible ripple carry adder achieves the improvement ratios ranging from 49.09\% to 54.09\%, and 13.40\% to 15.35\% in terms of delay compared to the designs presented in  \cite{Takahashi2005} and \cite{Takahashi_ripple}, respectively, while the design presented in \cite{Draper_ripple} is faster than the proposed design by 4.7\% to 9.02\%.

\begin{table*}[!ht]
\centering{
\small{
\caption{ A comparison of reversible ripple carry adder with no input carry }
\label{tabSR2}
\begin{tabular}{|l|l|l|l|l|}
\hline
 & 1 & 2 & 3 & \small{Proposed}  \\
\hline
 Ancilla Inputs & 1 & 0 & 0 & 0 \\
\hline
 Garbage Outputs &1 & 0 & 0 & 0\\
\hline
Quantum Cost& 17n-12 & 26n-29 & 15n-9 & 13n-8  \\ 
\hline
Delay $\Delta$ &10n &24n-27 & 13n-7  & 11n-4   \\
\hline
\multicolumn{5}{|l|}{ \scriptsize{ 1 is the design in \cite{Draper_ripple}}}\\
\multicolumn{5}{|l|}{ \scriptsize{ 2 is the design in \cite{Takahashi2005}}}\\
\multicolumn{5}{|l|}{ \scriptsize{ 3 is the design in \cite{Takahashi_ripple}}}\\
\hline
\end{tabular}
}
}
\end{table*}

\begin{table*}[!ht]
\centering{
\small{
\caption{ Quantum cost comparison of reversible ripple carry adders (no input carry)}
\label{tabSR2QC}
\begin{tabular}{|l|l|l|l|l|p{1.4cm}|p{1.4cm}|p{1.4cm}|}
\hline
Bits & 1 & 2 & 3 &\small{Proposed} & \% Impr. w.r.t 1 & \% Impr. w.r.t 2  & \% Impr. w.r.t 3 \\
\hline
 8 &  124& 179 & 111 & 96 & 22.5 &46.36&13.51 \\
\hline
 16 & 260& 387 & 231 & 200& 23  &48.32&13.41\\
\hline
32 & 532& 803 & 471 & 408 & 23.3 & 49.19&13.37 \\ 
\hline
64 & 1076& 1635 & 951 & 824 & 23.42 & 49.6& 13.35  \\ 
\hline
128 & 2164& 3299 & 1911 & 1656 &23.47 &49.8& 13.34  \\ 
\hline
256 & 4340& 6627 & 3831 & 3320 &23.5&  49.9&13.33  \\ 
\hline
512 & 8692& 13283 & 7671 & 6648 &23.51& 49.95& 13.33  \\ 
\hline
\multicolumn{8}{|l|}{ \scriptsize{ 1 is the design in \cite{Draper_ripple}}}\\
\multicolumn{8}{|l|}{ \scriptsize{ 2 is the design in \cite{Takahashi2005}}}\\
\multicolumn{8}{|l|}{ \scriptsize{ 3 is the design in \cite{Takahashi_ripple}}}\\
\hline
\end{tabular}
}
}
\end{table*}

\begin{table*}[!ht]
\centering{
\small{
\caption{ Delay(in $\Delta$) comparison of reversible ripple carry adders (no input carry)}
\label{tabSR2D}
\begin{tabular}{|l|l|l|l|l|p{1.4cm}|p{1.4cm}|p{1.4cm}|}
\hline
Bits & 1 & 2 & 3& \small{Proposed} &\% Impr. w.r.t 1 &\% Impr. w.r.t 2  & \% Impr. w.r.t 3  \\
\hline
 8 & 80 & 165 & 97 & 84&-  &49.09 &13.40\\
\hline
 16 &160& 357 & 201 &172&-  &51.8 &14.42\\
\hline
32 & 320& 741 & 409 & 348 & -&53.03 & 14.91 \\ 
\hline
64 & 640& 1509 &825 & 700 &-& 53.61 & 15.15 \\ 
\hline
128 & 1280& 3045 & 1657 & 1404 &-& 53.89 &15.26 \\ 
\hline
256 & 2560& 6117 & 3321 & 2812 &-& 54.02 & 15.32 \\ 
\hline
512 & 5120& 12261 & 6649 & 5628 &-& 54.09 &15.35 \\ 
\hline
\multicolumn{8}{|l|}{ \scriptsize{ 1 is the design in \cite{Draper_ripple}}}\\
\multicolumn{8}{|l|}{ \scriptsize{ 2 is the design in \cite{Takahashi2005}}}\\
\multicolumn{8}{|l|}{ \scriptsize{ 3 is the design in \cite{Takahashi_ripple}}}\\
\hline
\end{tabular}
}
}
\end{table*}

\section{ Design Methodology of Proposed Reversible ripple carry adder with input carry}
The reversible ripple carry adder with  input carry($c_0$) is designed without any ancilla inputs and the garbage outputs, and with less quantum cost and reduced delay compared to the existing design approaches which have optimized the adder design in terms of number of ancilla inputs. Consider the addition of two $n$ bit numbers  $a_{i}$ and $b_{i}$ stored at memory locations $A_{i}$ and $B_{i}$, respectively, where  $0 \le i \le n-1$.  The input carry $c_0$ is stored at memory location $A_{-1}$. Further, consider that memory location  $A_{n}$ is initialized with z $\in$ \{0, 1\}. At the end of the computation, the memory location $B_{i}$ will have  $s_{i}$, while the location $A_{i}$ keeps the value $a_{i}$ for  $0 \le i \le n-1$. Further, at the end of the computation, the additional location $A_{n}$ that initially stores the value z  will  have the value  $z\oplus s_{n}$, and the memory location $A_{-1}$ keeps the input carry $c_0$. Thus $A_{n}$  will have the value of $s_{n}$ when z=0.   Here, $s_{i}$ is the sum bit produced and is defined as: \\
$
     s_{i} = 
\begin{cases} 
 a_{i}\oplus b_{i} \oplus c_{i} & \mbox{if  $0\le i\le n-1$} \\
c_{n} & \mbox{if $i=n$} 
\end{cases}
$\\
where $c_{i}$ is the carry bit and is defined as:\\
$
     c_{i} = 
\begin{cases} 
 c_0 & \mbox{if } i=0 \\
 a_{i-1} b_{i-1} \oplus b_{i-1} c_{i-1} \oplus c_{i-1} a_{i-1} & \mbox{if } 1\le i\le n
\end{cases}
$
\\\\
\emph{
As shown above $c_{i}$ is the carry bit and is generated by using  $a_{i-1}$, $b_{i-1}$ and $c_{i-1}$. In our proposed approach firstly all the carry bits are generated and are saved in the memory location $A_{i-1}$ which was initially used for storing $a_{i-1}$  for $1\le i\le n-1$. Once the generated carry bits are used, the location $A_{i}$ are restored to the value $a_{i}$ while the location $B_{i}$ will have  $s_{i}$ for  $0 \le i \le n-1$. Thus restoring of location $A_{i}$ to the value $a_{i}$ helps in minimizing the garbage outputs. Since no constant input having the value as 0 is needed in the proposed approach, it saves the ancilla inputs.   The details of the proposed approach to minimize the garbage outputs and the ancilla inputs can be understood  by following the steps of the proposed design methodology.}
\begin{figure*}[!ht]
 \begin{center}
 \subfigure[After Step 1]
   {\label{8bitadder1ancilla1-1}
   \includegraphics[width=0.80in]{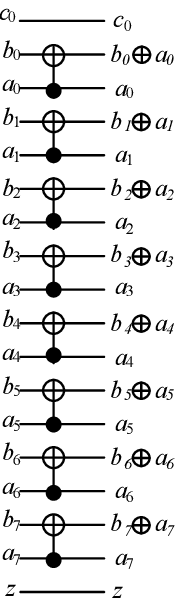}}
   \quad
\subfigure[After Step 2]
   {\label{8bitadder1ancilla1-2}
   \includegraphics[width=2.0in]{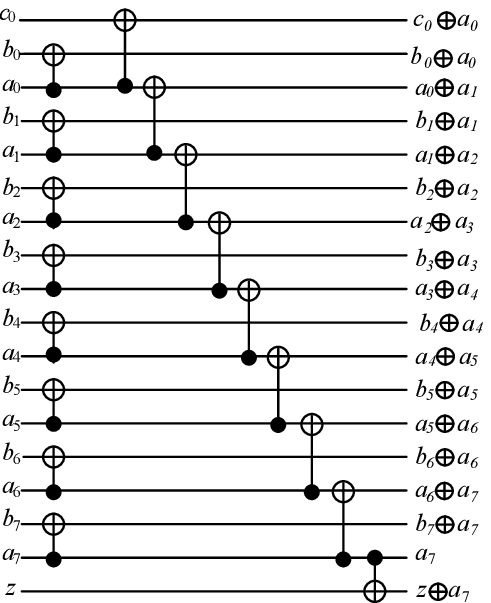}}
   \quad
       \subfigure[After Step 3]
   {\label{8bitadder1ancilla1-3}
   \includegraphics[width=2.85in]{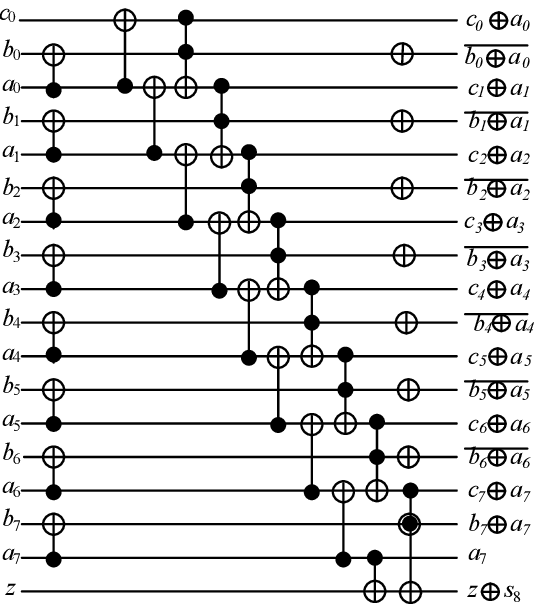}}
   \quad

 \end{center}
 \caption{Circuit generation of reversible 8 bit adder with input carry: Steps 1-3}
 \label{8bitadder1ancillainput1-3}
\end{figure*}

\begin{figure*}[!ht]
 \begin{center}
         \subfigure[After Step 4]
   {\label{8bitadder1ancilla1-4}
   \includegraphics[width=3.2in]{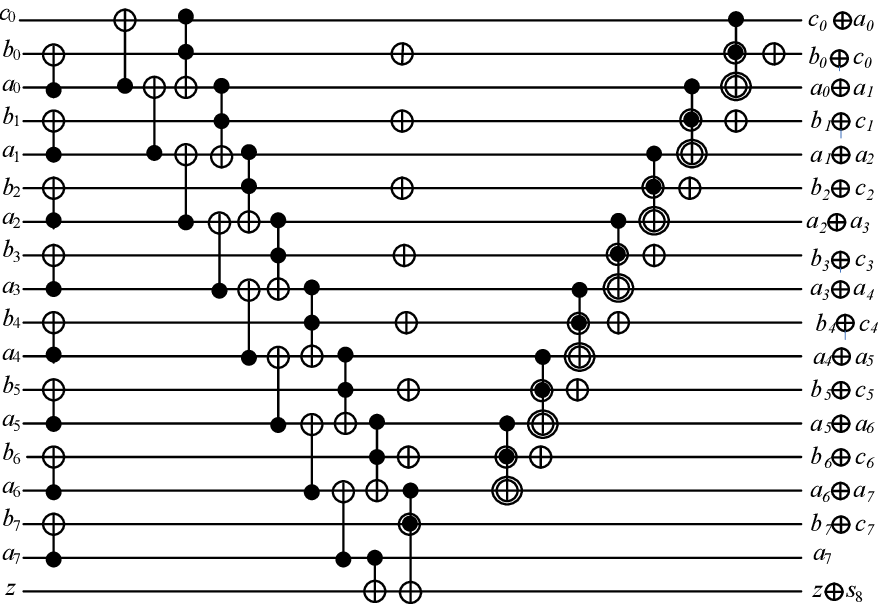}}
   \quad
       \subfigure[After Step 5]
   {\label{8bitadder1ancilla1-5}
   \includegraphics[width=3.3in]{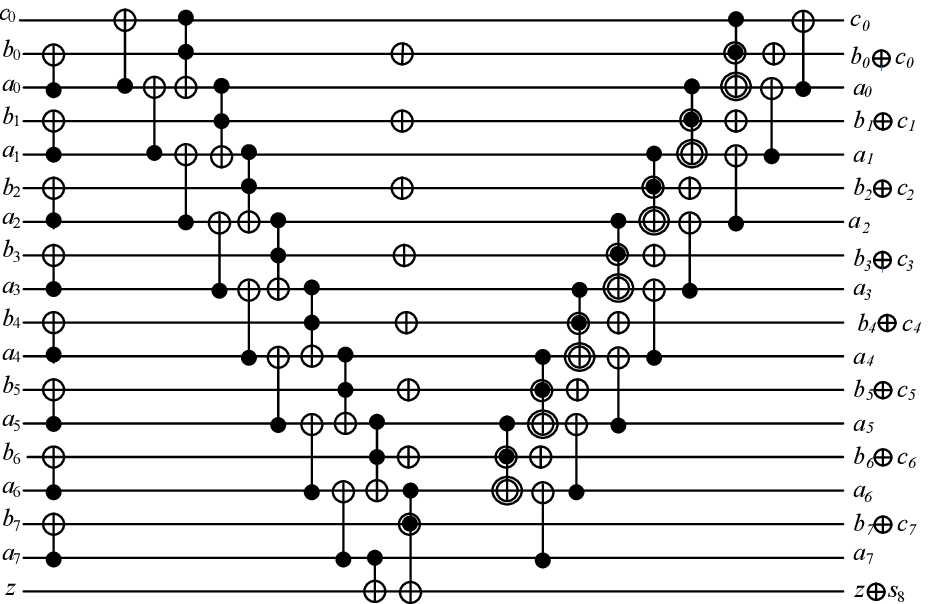}}
   \quad  
\subfigure[After Step 6]
   {\label{8bitadder1ancilla1-6}
   \includegraphics[width=3.3in]{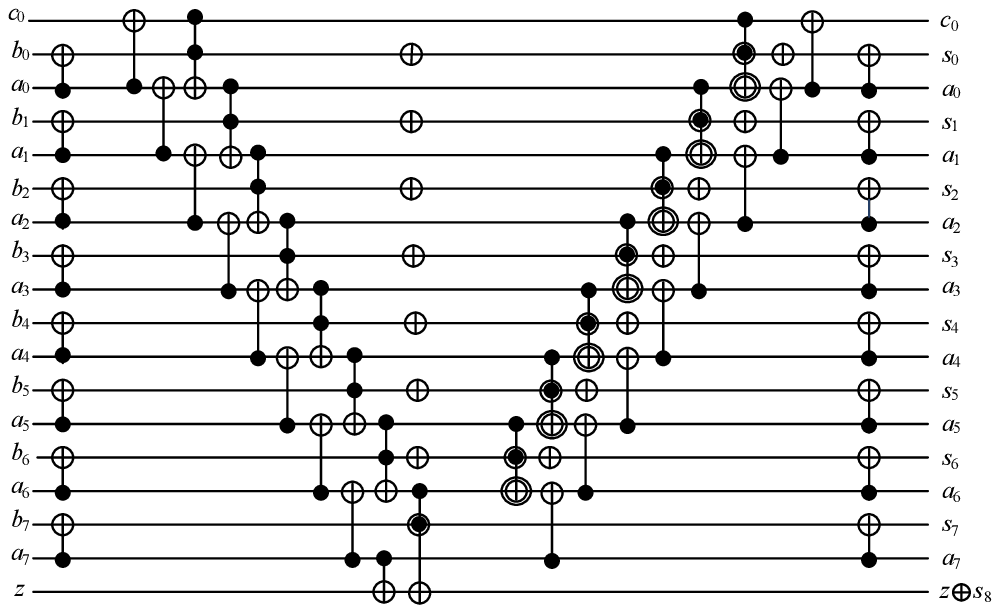}}
   \quad
      
 \end{center}
 \caption{Circuit generation of reversible 8 bit adder with input carry: Steps 4-6}
 \label{8bitadder1ancilla14-6}
\end{figure*}

The proposed method improves the delay and the quantum cost by selectively using the Peres gate and the TR gate at the appropriate places. The generalized methodology of designing the n bit reversible ripple carry adder with input carry is explained below along with an illustrative example of 8 bit reversible ripple carry adder.  The illustrative example of  8 bit reversible ripple carry adder is shown  in Figs. \ref{8bitadder1ancillainput1-3}  and \ref{8bitadder1ancilla14-6} that can perform the addition of two 8 bit numbers $a$=$a_0...a_7$ and $b$=$b_0...b_7$, and has the input carry $c_{0}$. The proposed methodology will be referred as methodology 2 further in this work and is explained in the following steps:\\

\textsf{Steps of Proposed Methodology 2 (Reversible Adder With Input Carry)}

\begin{enumerate}
\item For i=0 to n-1:\\ 
At pair of locations $A_i$ and $B_i$ apply the CNOT gate such that the location $A_i$ will maintain the same value, while location $B_i$ transforms to the value  *$A_{i}$ $\oplus$ *$B_{i}$, where *$A_{i}$ and *$B_{i}$ represent the values stored at location $A_{i}$ and $B_{i}$.  For illustrative purpose, the circuit of reversible ripple carry adder with input carry $c_{0}$ after step 1 is shown for addition of 8 bit numbers in Fig.\ref{8bitadder1ancilla1-1}. 

\item  For i= -1 to n-2: \\
At pair of locations $A_{i+1}$ and $A_{i}$ apply the CNOT gate such that the location $A_{i+1}$ will maintain the same value, while the value at location  $A_{i}$ transforms to  *$A_{i+1}$ $\oplus$ *$A_{i}$.  Further, apply a CNOT gate at pair of locations $A_{n-1}$ and $A_{n}$ such that the value at location $A_{n-1}$ will remain same, while the value at location  $A_{n}$ transforms to  *$A_{n-1}$ $\oplus$ *$A_{n}$.  The reversible 8 bit adder circuit after the  step 2 is illustrated in Fig. \ref{8bitadder1ancilla1-2}. 

\item  The step 3 has the following sub-steps:
  \begin{enumerate}
   \item For i=0 to n-2: \\
At locations $A_{i-1}$, $B_i$ and $A_i$ apply the Toffoli gate such that $A_{i-1}$, $B_i$ and  $A_{i}$ are passed to the inputs A, B, C, respectively, of the Toffoli gate.  Apply a Peres gate at location $A_{n-2}$, $B_{n-1}$ and $A_n$ such that  $A_{n-2}$, $B_{n-1}$ and  $A_{n}$ are passed to the inputs A, B, C, respectively, of the Peres gate. 
\item For i=0 to n-2: Apply a NOT gate at location $B_i$. 
\end{enumerate} 
The reversible 8 bit adder circuit based on the proposed design methodology after the step 3 is illustrated in  Fig. \ref{8bitadder1ancilla1-3}
\item  The step 4 has the following two sub-steps:
  \begin{enumerate}
    \item For i=n-2 to 0: \\
At locations $A_{i-1}$, $B_i$ and $A_{i}$ apply the TR gate such that $A_{i-1}$, $B_i$ and  $A_{i}$ are passed to the inputs A, B, C, respectively, of the TR gate.   

\item For i=0 to n-2:  Apply a NOT gate at location $B_i$.  
    \end{enumerate} 
The reversible 8 bit adder circuit based on the proposed methodology after step 4 is illustrated in Fig. \ref{8bitadder1ancilla1-4}. 
\item  For i=n-1 to 0: \\
At pair of locations $A_i$ and $A_{i-1}$ apply the CNOT gate such that the location $A_i$ will maintain the same value, while value at location $A_{i-1}$ transforms to the value  *$A_{i}$ $\oplus$ *$A_{i-1}$.  The reversible 8 bit adder circuit after the step 5 is shown in Fig.\ref{8bitadder1ancilla1-5}.

\item For i=0 to n-1: \\
At pair of locations $A_i$ and $B_{i}$ apply the CNOT gate such that the location $A_i$ will maintain the same value, while the value at location $B_i$ transforms to the value  *$A_{i}$ $\oplus$ *$B_{i}$.  After this step we will have the complete working design of the reversible adder an example of which is shown for addition of 8 bit numbers in Fig.\ref{8bitadder1ancilla1-6}.  
\end{enumerate}

\begin{figure}[!ht]
\begin{center}
\includegraphics[width=3.0in]{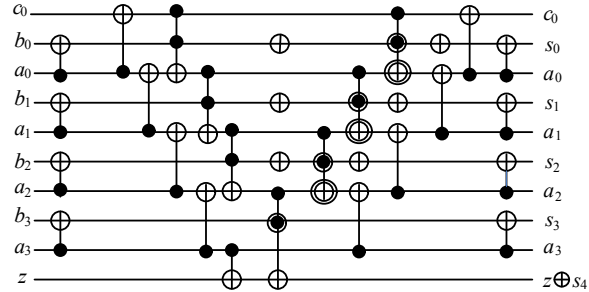}
\end{center} 
\caption{Proposed reversible 4 bit adder with input carry}
\label{4bitadder1ancilla}
\end{figure}

Thus, the proposed methodology is able to design the reversible ripple carry adder with an input carry without any ancilla and garbage bits. As in the design of the reversible BCD adder, the 4 bit reversible ripple carry adder will be used, thus its design is also illustrated in Fig.\ref{4bitadder1ancilla}\\ 

\textsf{Theorem 2}: \emph{Let a and b are two n bit binary numbers represented as $a_{i}$ and $b_{i}$, $c_0$ is the input carry ($c_0$), and  z $\in$ \{0, 1\} is the another 1 bit input, where $0 \le i \le n-1$,  then the proposed design steps of methodology 2 result in the ripple carry adder circuit that works correctly. The proposed design methodology designs an n bit adder circuit that  produces the sum output $s_{i}$ at the memory location where $b_{i}$ is initially stored, while the location where  $a_{i}$ is initially stored  is restored to the value $a_{i}$ for  $0 \le i \le n-1$. Further, the memory location where z is initially stored transforms to  $z\oplus s_{n}$, and  the memory  location where the input carry $c_0$ is initially stored is restored to the value $c_0$}.\\\\
\textsf{Proof}:  The proposed approach will make the following changes on the inputs that are illustrated as follows:

\begin{enumerate}
\item Step 1: The step 1 of the proposed approach transforms the input states to\\
$\left |c_0 \right \rangle   \left (\bigotimes _{i=0}^{n-1} \left | b_i \oplus a_i \right \rangle  \left | a_i \right \rangle   \right )  \left | z  \right \rangle$ \\

For illustrative purpose, the transformation of the input states of a 8 bit reversible adder circuit after step 1 is shown in Fig.\ref{8bitadder1ancilla1-1}.

\item Step 2: The step 2 of the proposed approach transforms the input states to\\
\begin{eqnarray*}
\left | c_0 \oplus a_0 \right \rangle \left (\bigotimes _{i=0}^{n-2} \left | b_i \oplus a_i \right \rangle  \left | a_i \oplus a_{i+1}\right \rangle   \right )  \left | b_{n-1} \oplus a_{n-1} \right \rangle \left | a_{n-1} \right \rangle \left   | z \oplus a_{n-1} \right \rangle 
\end{eqnarray*} 

For illustrative purpose, the transformation of the input states of a 8 bit reversible adder circuit after step 2 is shown in Fig.\ref{8bitadder1ancilla1-2}.

\item Step 3: The step 3 has the two sub-steps:\\
Step 3.a has the Toffoli gates which take the inputs as  $c_i \oplus a_i$,  $b_i \oplus a_i$,  and $a_i \oplus a_{i+1}$ for  $0 \le i \le n-2$. The Toffoli gates will produce the outputs as $c_i \oplus a_i$,  $b_i \oplus a_i$,  and $c_{i+1} \oplus a_{i+1}$.  The third outputs of the Toffoli gates are  $c_{i+1} \oplus a_{i+1}$  because of the fact that the Toffoli gate has the logic equation as  $A\cdot B \oplus C$ where A, B and C are the inputs of a Toffoli gate, thus will produce the output as 
$c_i \oplus a_i \cdot  b_i \oplus a_i \oplus  a_i \oplus a_{i+1}$= $c_{i+1} \oplus a_{i+1}$. Finally we have a  Peres gate having the inputs $c_{n-1} \oplus a_{n-1}$,  $b_{n-1} \oplus a_{n-1}$,  and $a_{n-1} \oplus z$ which produces the outputs as $c_{n-1} \oplus a_{n-1}$,  $c_{n-1} \oplus b_{n-1}$,  and $s_{n} \oplus z$. Thus after the step 3.a the input states are transformed to:
\begin{eqnarray*}
\left | c_0 \oplus a_0 \right \rangle \left (\bigotimes _{i=0}^{n-2} \left | b_i \oplus a_i \right \rangle  \left | c_{i+1} \oplus a_{i+1}\right \rangle   \right )  \left | b_{n-1} \oplus c_{n-1} \right \rangle \left | a_{n-1} \right \rangle \left   | z \oplus s_{n} \right \rangle 
\end{eqnarray*}

The step 3.b applies the NOT operation to the location $B_i$ having the value as $b_i \oplus a_i$ for $0 \le i \le n-2$, thus the input states are transformed to

\begin{eqnarray*}
\left | c_0 \oplus a_0 \right \rangle \left (\bigotimes _{i=0}^{n-2} \overline{ \left | b_i \oplus a_i \right \rangle}  \left | c_{i+1} \oplus a_{i+1}\right \rangle   \right )  \left | b_{n-1} \oplus c_{n-1} \right \rangle  \left | a_{n-1} \right \rangle \left   | z \oplus s_{n} \right \rangle 
\end{eqnarray*} \\\\

For illustrative purpose, the transformation of the input states of a 8 bit reversible adder circuit after step 3 is shown in Fig.\ref{8bitadder1ancilla1-3}.

\item Step 4: The step 4 has the TR gates which take the inputs as  $\{c_{i} \oplus a_{i}$,  $\overline{b_i \oplus a_i}$,  and $c_{i+1} \oplus a_{i+1}$ for i=n-2 to 0. The TR gates will produce the outputs as $c_i \oplus a_i$,  $\overline{b_i \oplus c_i}$,  and $a_{i} \oplus a_{i+1}$ for i=n-2 to 0. Thus after the application of TR gates the input states transform to: 

\begin{eqnarray*} 
\left | c_0 \oplus a_0 \right \rangle \left (\bigotimes _{i=0}^{n-2} \left | \overline{b_i \oplus c_i }\right \rangle  \left | a_{i} \oplus a_{i+1}\right \rangle   \right )  \left | b_{n-1} \oplus c_{n-1} \right \rangle  \left | a_{n-1} \right \rangle \left   | z \oplus s_{n} \right \rangle
\end{eqnarray*}

Next, the NOT gates are applied to the TR gates outputs $\overline{b_i \oplus c_i}$ for i=n-2 to 1. Thus the step 4 of the proposed approach transforms the input states to\\

\begin{eqnarray*} 
\left | c_0 \oplus a_0 \right \rangle \left (\bigotimes _{i=0}^{n-2} \left | b_i \oplus c_i \right \rangle  \left | a_{i} \oplus a_{i+1}\right \rangle   \right )  \left | b_{n-1} \oplus c_{n-1} \right \rangle  \left | a_{n-1} \right \rangle \left   | z \oplus s_{n} \right \rangle
\end{eqnarray*}

For illustrative purpose, the transformation of the input states of a 8 bit reversible adder circuit after step 4 is shown in Fig.\ref{8bitadder1ancilla1-4}.

\item Step 5: The step 5 of the proposed approach transforms the input states to\\
\begin{eqnarray*}
\left | c_0 \right \rangle \left (\bigotimes _{i=0}^{n-2} \left | b_i \oplus c_i \right \rangle  \left | a_{i} \right \rangle   \right )  \left | b_{n-1} \oplus c_{n-1} \right \rangle  \left | a_{n-1} \right \rangle \left   | z \oplus s_{n} \right \rangle 
\end{eqnarray*}

For illustrative purpose, the transformation of the input states of a 8 bit reversible adder circuit after step 5 is shown in Fig.\ref{8bitadder1ancilla1-5}.

\item Step 6: The step 6 of the proposed approach transforms the input states to\\
\begin{eqnarray*}
\left | c_0 \right \rangle \left (\bigotimes _{i=0}^{n-1} \left | s_i \right \rangle  \left | a_{i} \right \rangle   \right )  \left   | z \oplus s_{n} \right \rangle 
\end{eqnarray*}

For illustrative purpose, the transformation of the input states of a 8 bit reversible adder circuit after step 6 is shown in Fig.\ref{8bitadder1ancilla1-6}.

\end{enumerate}

Thus we can see that the proposed six step will produce the sum output $s_{i}$ at the memory location where $b_{i}$ is stored initially, while the location where  $a_{i}$ is stored initially will be restored to the value $a_{i}$ for  $0 \le i \le n-1$. The memory location where z is stored will have  $z\oplus s_{n}$ and  the memory  location where the input carry $c_0$ was stored initially will be restored to the value $c_0$.This proves the correctness of the proposed methodology of designing the reversible ripple carry adder with input carry.\\\\

\emph{ Delay and Quantum Cost}
\begin{itemize}
\item Step 1 of the proposed methodology needs $n$ CNOT gates working in parallel thus this step has the quantum cost of $n$ and delay of $1 \Delta$.
\item Step 2 of the proposed methodology needs $n+1$ CNOT gates working in series thus this step has the quantum cost of $n+1$. The delay of this stage will be only $2 \Delta$ as it has $n-1$ CNOT gates work in parallel with the Toffoli gates of the next stage thus only $2$ CNOT gates contributes to the delay..
\item  Step 3 needs $n-1$ Toffoli gates working in series thus  contributing to the quantum cost of $5(n-1)$ and delay of $5(n-1)$ $\Delta$. There is a Peres gate contributing to the quantum cost of $4$ and delay of $4  \Delta$. There are $n-1$ NOT gates working in parallel with the Peres gate thus contributing to quantum cost of $n-1$ and zero delay.  The total quantum cost of this stage is $5(n-1)+4+n-1$ while the delay contribution of this stage is $5(n-1) \Delta+4 \Delta$. 

\item  Step 4 needs $n-1$ TR gates working in series thus contributing to  the quantum cost by $4(n-1)$ and delay of $4(n-1) \Delta$. Further, there are $n-1$ NOT gates, which all work in parallel with the TR gates except the last NOT gate. Thus, it contributes to quantum cost of $n-1$ and delay of $1 \Delta$. Thus this step has the quantum cost of  $4(n-1)+n-1$ and the delay of $4(n-1) \Delta$ +1 $\Delta$. 
\item  Step 5 needs $n$ CNOT gates working in parallel with the TR gates and the NOT gates, except the last one.  Thus this step has the quantum cost of $n$ and delay of $1 \Delta$.
\item  Step 6 needs $n$ CNOT gates working in parallel thus this step has the quantum cost of $n$ and delay of $1 \Delta$.
\end{itemize}
Thus the total quantum cost of n bit reversible ripple carry adder is $n+n+1+5(n-1)+4+n-1+4(n-1)+n-1+n+n = 15n-6$. The propagation delay will be $ 1 \Delta+2 \Delta+5(n-1) \Delta +4 \Delta+4(n-1) \Delta+1 \Delta+1 \Delta+1 \Delta = (9n+1) \Delta$.

A comparison of the proposed design with the existing designs is illustrated in Table \ref{tabSR21} which shows that the proposed design of reversible ripple carry adder with input carry is designed with no ancilla input bit and has less quantum cost and delay compared to its existing counterparts.
Table \ref{tabSR21QC} shows the comparison in terms of quantum cost which shows that the proposed design of the reversible carry adder with  input carry achieves the improvement ratios ranging from 12.3\% to 11.77\% and 0\% to 11.61\% compared to the designs presented in \cite{Draper_ripple}. From Table \ref{tabSR21D}, it can be seen that the proposed design of reversible ripple carry adder achieves the improvement ratios ranging from  10.97\% to 10.01\%, and 0\% to 9.83\% in terms of delay compared to the designs presented in  \cite{Draper_ripple}, respectively.

\begin{table*}[!ht]
\centering{
\small{
\caption{ A comparison of reversible ripple carry adder with  input carry }
\label{tabSR21}
\begin{tabular}{|c|c|c|c|}
\hline
 &  1  & 2& Proposed   \\
\hline
 Ancilla Inputs & 0 & 0 & 0 \\
\hline
 Garbage Outputs &0 & 0 & 0 \\
\hline
Quantum Cost& 17n-6 & 17n-22 & 15n-6  \\ 
\hline
Delay $\Delta$ &10n+2 &10n-8 & 9n+1   \\
\hline
\multicolumn{4}{|l|}{ \scriptsize{ 1 is the design 1 in \cite{Draper_ripple}}}\\
\multicolumn{4}{|l|}{ \scriptsize{ 2 is the design 2 in \cite{Draper_ripple}}}\\
 
\hline
\end{tabular}
}
}
\end{table*}

\begin{table*}[!ht]
\centering{
\small{
\caption{ Quantum cost comparison of reversible ripple carry adders (with input carry)}
\label{tabSR21QC}
\begin{tabular}{|l|l|l|l|p{1.4cm}|p{1.4cm}|}
\hline
Bits & 1 & 2 & \small{Proposed} &\% Impr. w.r.t 1  & \% Impr. w.r.t 2  \\
\hline
 8 & 130 & 114 & 114  &12.30&- \\
\hline
 16 &266 & 250 & 234  &12.03&6.4\\
\hline
32 & 538 & 522 & 474 & 11.89&9.19 \\ 
\hline
64 & 1082 & 1066 & 954 &11.82& 10.50  \\ 
\hline
128 & 2179 & 2154 & 1914 &11.79& 11.14 \\ 
\hline
256 & 4346 & 4330 & 3834 & 11.78&11.45  \\ 
\hline
512 & 8698 & 8682 & 7674 & 11.77& 11.61  \\ 
\hline
\multicolumn{6}{|l|}{ \scriptsize{ 1 is the design 1 in \cite{Draper_ripple}}}\\
\multicolumn{6}{|l|}{ \scriptsize{ 2 is the design 2 in \cite{Draper_ripple}}}\\
\hline
\end{tabular}
}
}
\end{table*}

\begin{table*}[!ht]
\centering{
\small{
\caption{ Delay (in $\Delta$) comparison of reversible ripple carry adders (with input carry)}
\label{tabSR21D}
\begin{tabular}{|l|l|l|l|p{1.4cm}|p{1.4cm}|}
\hline
Bits & 1 & 2 & \small{Proposed} &\% Impr. w.r.t 1  & \% Impr. w.r.t 2  \\
\hline
 8 & 82 & 72 & 73  &10.97&- \\
\hline
 16 &162 & 152 & 145  &10.49&4.6\\
\hline
32 & 322 & 312 & 289 & 10.24&7.37 \\ 
\hline
64 & 642 & 632 & 577 &10.12& 8.7  \\ 
\hline
128 & 1282 & 1272 & 1153 &10.06& 9.35 \\ 
\hline
256 & 2562 & 2552 & 2305 & 10.03& 9.67  \\ 
\hline
512 & 5122 & 5112 & 4609 & 10.01& 9.83  \\ 
\hline
\multicolumn{6}{|l|}{ \scriptsize{ 1 is the design 1 in \cite{Draper_ripple}}}\\
\multicolumn{6}{|l|}{ \scriptsize{ 2 is the design 2 in \cite{Draper_ripple}}}\\
\hline
\end{tabular}
}
}
\end{table*}

\section{Design of Reversible BCD Adder}
A  BCD adder is a circuit that adds two BCD digits in parallel and produces a sum digit, also in BCD.  We are illustrating two different approaches of designing the conventional BCD adder. 

\subsection{Basics}
\begin{figure}[!ht]
 \begin{center}
 \subfigure[Approach 1 for design of conventional 1 digit BCD adder ]
   {\label{4bitbcdadder}
   \includegraphics[width=2.5in]{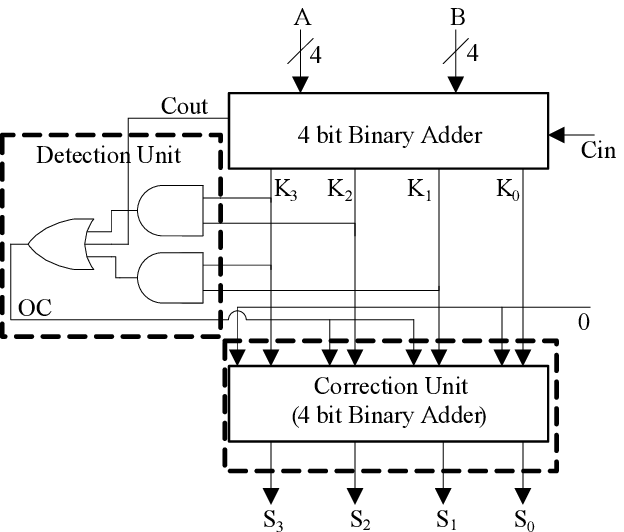}}
   \quad
\subfigure[Approach 2 for design of conventional 1 digit BCD adder]
   {\label{4bitbcdadder1}
   \includegraphics[width=2.0in]{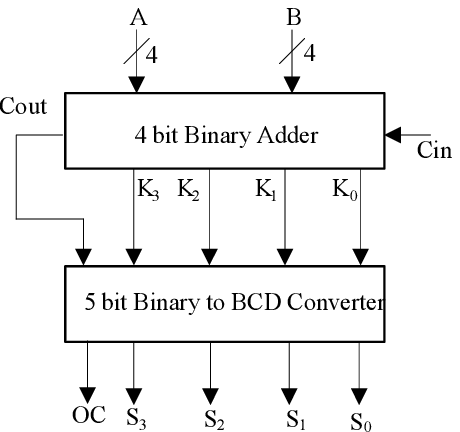}}
   \quad
      
 \end{center}
 \caption{ Design approaches of conventional BCD adder }
 \label{bcdbitadder}
\end{figure}

Figure \ref{4bitbcdadder} shows the  first approach of designing the 1 digit conventional BCD adder, which also includes the detection and the correction logic in its internal construction. The two decimal digits A and B, together with the input carry Cin, are first added in the top 4-bit binary adder to produce the 4 bit binary sum ($K_3$ to $K_0$) and the carry out (Cout). In the BCD addition, when the binary sum of A and B is less than 1001, the BCD number is same as the binary number thus no conversion is needed. But when the binary sum of A and B is greater than 1001, 0110 is added to convert the binary number into an equivalent BCD number. The condition that summation of numbers A, B and Cin is greater than 1001 is detected through a detection unit. The detection unit works on the condition that can be expressed by the Boolean function $OC=Cout+ K_3 \cdot K_2+K_3\cdot K_1$. When OC (output carry) is equal to zero, nothing is added to the binary sum. When it is equal to one, binary 0110 is added to the binary sum using the correction unit (another 4-bit binary adder).  

As illustrated above in Fig.\ref{4bitbcdadder}, the  binary adder produces a result that may not be in correct BCD format and need to be converted to BCD format through the use of detection and correction unit. Instead of using the detection and the correction unit to convert the result of summation to the BCD format, the outputs of the binary adder can be passed to a binary to BCD converter to have the result of the binary addition in the BCD format \cite{majid_bcd2009}. This approach is illustrated in the Fig. \ref{4bitbcdadder1} where the 5 bit binary to BCD converter produces the desired output in the BCD format. In this work, we have proposed the equivalent reversible design of these two approaches to design the reversible BCD adder optimized for the number of ancilla input bits and the number of garbage outputs. 

\subsection{ Design of Reversible BCD Adder Based on Approach 1}
We present two designs of reversible BCD adders based on approach 1 with and without input carry $c_0$ (the conventional irreversible design of the 1 digit BCD adder  based on approach 1 is illustrated in Fig. \ref{4bitbcdadder}).

\subsubsection{Design 1 of reversible BCD adder with input carry}
As can be observed from Fig.\ref{4bitbcdadder} that  in order to have this design, we need the design of 4 bit reversible adder with input carry, the design of which is already illustrated in Fig.\ref{4bitadder1ancilla}. Next we present the new reversible design of the detection unit. As illustrated above, the detection unit uses the Boolean function $OC=Cout+ K_3 \cdot K_2+K_3\cdot K_1$ as the checking condition. It can be written as $OC=Cout+ K_3( K_2+ K_1)$. On careful observation it can be reduced to $OC=Cout\oplus K_3( K_2+ K_1)$ as Cout and $K_3( K_2+ K_1)$ cannot be true at the same time \cite{Biswas}. We have designed the reversible detection unit based on the modified Boolean equation $OC=Cout\oplus K_3( K_2+ K_1)$ using NOT, CNOT, the Peres gate and the TR gate. Before explaining the reversible design of the detection unit, we would like to emphasize a very useful property of the TR gate in relation to the popular Peres gate. We derive the inverse of the TR gate since a reversible gate can be combined with its inverse reversible gate to minimize the garbage outputs \cite{Fredkin}. In order to derive the logic equations of the inverse TR gate, we performed the reverse mapping of the TR gate outputs working as inputs to generate the inputs of the TR gate. We observe that the inverse of the TR gate is same as the existing Peres gate having inputs to outputs mapping as (P=A, Q=A $\oplus$ B, $R=A \cdot B \oplus C$. Thus, the TR gate and the Peres gate are inverse of each other.

\begin{figure}[!ht]
 \begin{center}
 \subfigure[Proposed reversible detection unit with one ancilla input bit. $K_1, K_2,K_3$ and Cout here are the outputs of the top 4 bit binary adder needed in the design of 1 digit BCD adder. OC represents the output carry. The details of $K_1,K_2, K_3$, Cout and OC signals can be found in Fig.\ref{4bitbcdadder}]
   {\label{detection}
   \includegraphics[width=2.25in]{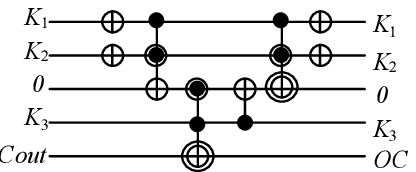}}
   \quad
\subfigure [Proposed reversible correction unit with no ancilla input bit. $K_0$ here represents the least significant sum bit generated by the 4 bit binary adder in the BCD adder. $K_1, K_2, K_3$ and OC are the outputs of the reversible detection unit. The details of these signals can also be found in Fig.\ref{4bitbcdadder} ]
   {\label{correction}
   \includegraphics[width=2.25in]{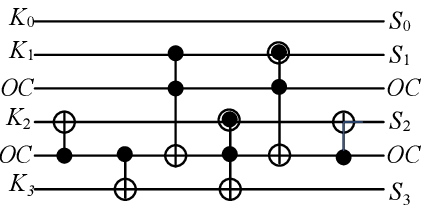}}
   \quad
 \end{center}
\caption{ Proposed detection and correction unit of reversible BCD adder}
 \label{detection_correction}
\end{figure}

     The design of the reversible detection unit is illustrated in Fig.\ref{detection} in which by using the TR gate as the inverse of the Peres gate we are able to regenerate $K_1,K_2,K_3$ to be used further in the correction unit. Further, the ancilla input bit having the constant value as '0' is regenerated to be used further in the correction unit. In the design, firstly with the help of the NOT gate and the Peres gate the output $\bar{K_1} \cdot \bar{K_2}$ is generated which is passed to the TR gate to generate the  output $OC=Cout\oplus K_3( K_2+ K_1)$. Then with the help of the CNOT gate cascaded at the inputs A and B of the TR gate, the function  $\bar{K_1} \cdot \bar{K_2}$ is regenerated. Finally, TR gate combined with NOT gates is used to regenerate  $K_1,K_2,0$ outputs (here the final TR gate works as the inverse of the Peres gate). The reversible detection unit has the quantum cost of 17 and the delay of 15 $\Delta$. 
 
\begin{figure}[!ht]
\begin{center}
\includegraphics[width=3.5in]{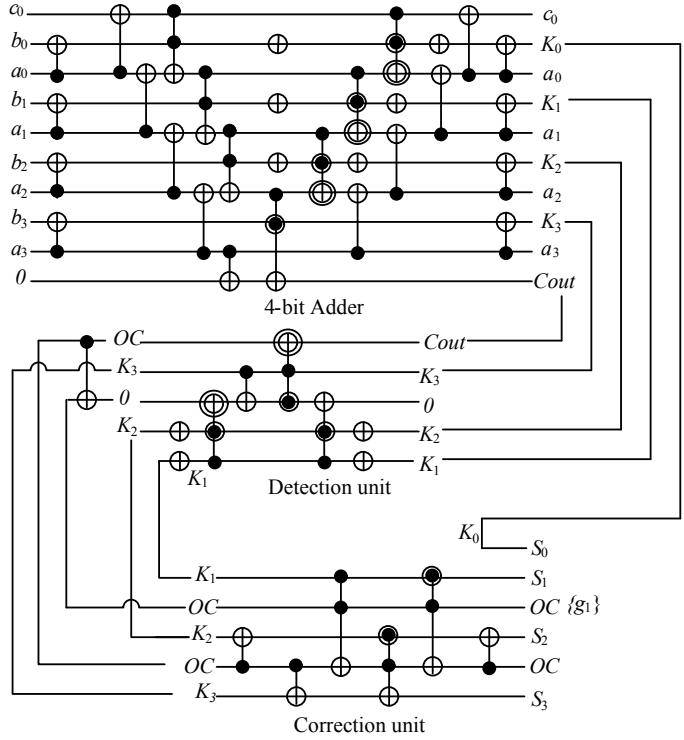}
\end{center} 
\caption{Proposed design of 1 digit reversible BCD adder  with input carry (RBCD-1) based on approach 1. The design consists of 4 bit reversible binary adder with input carry illustrated in Fig.\ref{4bitadder1ancilla}, reversible detection unit illustrated in Fig.\ref{detection} and reversible correction unit illustrated in Fig.\ref{correction}. $g_1$ which is the extra copy of the OC output represent the only garbage output. There are two ancilla inputs with constant value of 0. The copy of the OC signal is made through a Feynman gate as fanout is not allowed in reversible logic. }
\label{RBCDAdder1}
\end{figure}

    The design of the reversible correction unit is illustrated in Fig.\ref{correction}. The design of the correction unit is a 2 bit binary adder based on the methodology of the proposed reversible ripple carry adder  without input carry as illustrated  earlier in section IV. In the design the 2 bits inputs of the adders are {$a_0=K_1, b_0=OC$}, {$a_1=K2, b_1=OC$}. In order to generate $S_3$ we have passed OC at the location z ($a_3$) of the reversible ripple carry adder as S3 can be generated as $K_3 \oplus C_3$. Since 0 needs to be added to $K_0$ to produce  $S_0$, thus $S_0$ will be same as $K_0$ and hence a wire connection. The proposed design of the reversible correction unit does not need any ancilla input bit.  The reversible correction unit has the quantum cost of 16 and delay of 16 $\Delta$. The modules designed above can be integrated together to design the 1 digit reversible BCD adder as  illustrated in Fig.\ref{RBCDAdder1}. The design Fig.\ref{RBCDAdder1} will be used with name RBCD-1 further in this work. The proposed design contains the 4 bit reversible adder with input carry, reversible detection unit and reversible correction unit. A Feynman gate is used to avoid the fanout of OC signal as fanout is not allowed in reversible logic. It can be observed that the proposed reversible BCD adder design uses two ancilla input bits, and generates 1 garbage output labelled as g1 in the Fig.\ref{RBCDAdder1} which is the copy of the output carry (OC) that will not be used further in the computation \emph{(the inputs regenerated at the outputs are not considered as garbage outputs)}. The design has the quantum cost of 88 which is the summation of the quantum cost of 4 bit reversible adder with input carry, reversible detection unit, 1 Feynman gate and reversible correction unit. Further, the design has the delay of 73 $\Delta$ which is the summation of the propagation delay of the 4 bit reversible adder with input carry, reversible detection unit, 1 Feynman gate and reversible correction unit.
    
     \emph{Once we have designed the 1 digit reversible BCD adder, the $n$ digit reversible BCD adder with input carry can be designed by cascading of the 1 digit reversible BCD adder (RBCD-1) in the ripple carry fashion as illustrated in Fig.\ref{ndigitd1}}. Thus, the design 1 of  $n$ digit reversible BCD adder with input carry has  $2n$ ancilla inputs bits, $2n-1$ garbage outputs, quantum cost of $88n$ and delay of  $73n \Delta$. As shown in Fig.\ref{ndigitd1} the design 1 of the $n$ digit reversible BCD adder with input carry  has $2n-1$ garbage outputs because the first 1 digit BCD adder will have 1 garbage output(extra OC output), while the remaining $n-1$ 1 digit BCD adders each will have two garbage outputs. The extra one garbage output in each $n-1$ 1 digit reversible BCD adder is from the ripple carry regenerated at the outputs, for example the second 1 digit BCD adder in Fig.\ref{ndigitd1} has two garbage outputs  labeled as g2 and g3, the extra garbage output g2 is the  output carry OC1 of the first 1 digit BCD adder that is passed to the second 1 digit BCD adder as input carry and is regenerated at one of its outputs.

 Table \ref{tabSR2123}  illustrates that the proposed design is better than the existing design in terms of ancilla input bits and garbage outputs while also being efficient in terms of the quantum cost and the delay.

\subsubsection{Design 2 of reversible BCD adder with no input carry}
\begin{figure}[!ht]
\begin{center}
\includegraphics[width=3.5in]{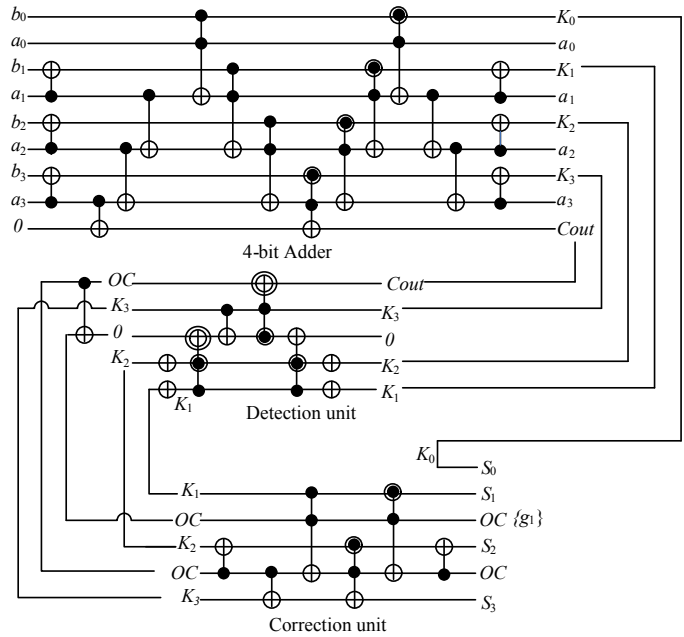}
\end{center}
\caption {Proposed design of 1 digit reversible BCD adder  with no input carry (RBCD-2) based on approach 1.  The design consists of 4 bit reversible binary adder without input carry illustrated in Fig.\ref{4bitadder0ancilla}, reversible detection unit illustrated in Fig.\ref{detection} and reversible correction unit illustrated in Fig.\ref{correction}. There are two ancilla inputs with constant value of 0 and $g_1$ represent the only garbage output. The copy of the OC signal is made through a Feynman gate as fanout is not allowed in reversible logic.}
 \label{RBCDAdder1noinputc}
  \end{figure}

\begin{figure*}[!ht]
 \begin{center}
\subfigure [Proposed design 1 of n digit reversible BCD adder with input carry]
   {\label{ndigitd1}
   \includegraphics[width=4.9in]{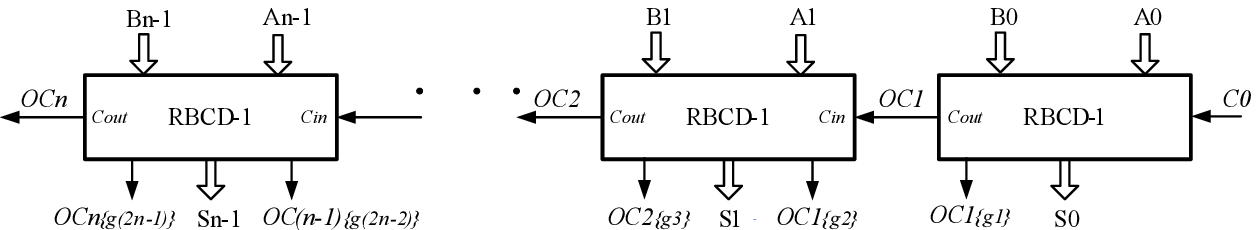}}
   \quad   
\subfigure [Proposed design 2 of n digit reversible BCD adder with no input carry]
   {\label{ndigitd2}
   \includegraphics[width=4.9in]{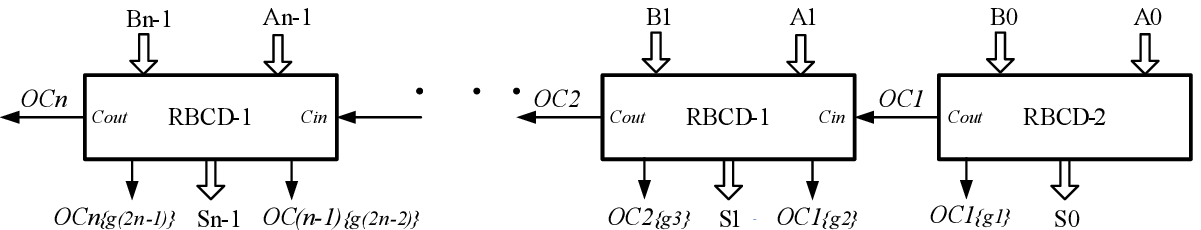}}
   \quad   
 \end{center}
 \caption{ Proposed designs of n digit reversible BCD adder based on approach 1}
 \label{BCDAdder12design}
\end{figure*}

  We have also designed 1 digit reversible BCD adder based on approach 1 having no input carry. For achieving the design efficient in terms of number of ancilla input bits and the garbage outputs, we have used the 4 bit reversible input adder without any input carry based on the methodology proposed in this work (the design can be referred in Fig. \ref{4bitadder0ancilla}).  The rest of the design is same as the design of the reversible BCD adder with input carry. The complete design of the 1 digit reversible BCD adder with no input carry  is illustrated in Fig.\ref{RBCDAdder1noinputc}, and will be used with name RBCD-2 further in this work. The design has only 2 ancilla input bits and needs 1 garbage output. The design has the quantum cost of 80 and delay of 80 $\Delta$. Once we have designed the 1 digit reversible BCD adder with no input carry (RBCD-2) , the $n$ digit reversible BCD adder with no input carry can be designed by utilizing the 1 digit reversible BCD adder with no input carry (RBCD-2) to add the least significant digit and then cascading  $n-1$  1 digit reversible BCD adder with input carry (RBCD-1) in the ripple carry fashion. The design of n digit reversible BCD adder with no input carry is illustrated in Fig.\ref{ndigitd2}. Thus, the design 2 of $n$ digit reversible BCD adder without input carry has  $2n$ ancilla inputs bits, $2n-1$ garbage outputs, quantum cost of $88n-18$ and delay of  $73n-1 \Delta$. Table \ref{tabSR2123}  illustrates that the proposed design is better than the existing design in terms of number of ancilla input bits and the garbage outputs while also being efficient in terms of the quantum cost and the delay.

\begin{table*}[!ht]
\centering{
\small{
\caption{ A Comparison of n digit reversible BCD adders}
\label{tabSR2123}
\begin{tabular}{|p{6cm}|p{1cm}|p{1.2cm}|p{1.2cm}|p{1cm}|}
\hline
 & Ancilla input & Garbage outputs & Quantum cost & Delay $\Delta$\\
\hline
 \cite{babu_bcd06} & 17n &18n & 110n & - \\
\hline
 \cite{Biswas} & 7n & 6n & 55n & - \\
\hline
 \cite{Thomsen} & 4n & 4n & 169n & - \\
\hline
 \cite{Majid_bcd1} & 14n & 16n & 84n & - \\
\hline
\cite{majid_bcd2009} (Design 3$^*$) & 2n & 6n & 103n & - \\
\hline
 This work 1 (with input carry) & 2n & 2n-1 & 88n  & 73n \\
\hline
 This work 2 (without input carry) & 2n & 2n-1 & 88n-18  & 73n-1 \\
\hline
This work 3 (with input carry) & n & n-1 & 70n  & 57n \\
\hline
This work 4 (without input carry) & n & n-1 & 70n-8  & 57n-3 \\
\hline
\hline 
\multicolumn{5}{|p{12cm}|}{ \scriptsize{ 1 represents proposed Design 1 of Reversible BCD adder with input carry}}\\
\multicolumn{5}{|p{12cm}|}{ \scriptsize{ 2 represents proposed Design 2 of Reversible BCD adder without input carry}}\\
\multicolumn{5}{|p{12cm}|}{ \scriptsize{ 3 represents proposed Design 3 of Reversible BCD adder without input carry}}\\
\multicolumn{5}{|p{12cm}|}{ \scriptsize{ 4 represents proposed Design 4 of Reversible BCD adder without input carry}}\\
\multicolumn{5}{|p{12cm}|} { \scriptsize{ $^*$In \cite{majid_bcd2009}, 6 designs of the BCD adders are proposed varying in parameters of the number of ancilla inputs, garbage outputs, quantum cost and the delay. Among 6 designs, design 3 has the minimum number of ancilla inputs and the garbage outputs, thus we have compared to our work with design 3 of the \cite{majid_bcd2009}.}}\\
\hline
\end{tabular}
}
}
\end{table*}

\begin{table*}[!ht]
\centering{
\small{
\caption{ Ancilla inputs comparison of n digit reversible BCD adders}
\label{tabSR21DANC}
\begin{tabular}{|l|l|l|l|p{1.4cm}|p{1.4cm}|}
\hline
Digits & 1 & 2 & \small{Proposed*} &\% Impr. w.r.t 1  & \% Impr. w.r.t 2  \\
\hline
 8 & 32 & 16 & 8  &75& 50 \\
\hline
 16 &64 & 32 & 16  &75&50\\
\hline
32 & 128 & 64 & 32 & 75&50 \\ 
\hline
64 & 256 & 128 & 64 &75 &50  \\ 
\hline
128 & 512 & 256 & 128 &75& 50 \\ 
\hline
256 & 1024 & 512 & 256 &75& 50  \\ 
\hline
512 & 2048 & 1024 & 512 & 75& 50  \\ 
\hline
\multicolumn{6}{|l|}{ \scriptsize{ 1 is the design  in \cite{Thomsen}}}\\
\multicolumn{6}{|l|}{ \scriptsize{ 2 is the design 3 in \cite{majid_bcd2009}}}\\
\multicolumn{6}{|p{8cm}|}{ \scriptsize{ *  is  our  design 3 proposed in this work. In improvement calculation all the existing  designs are compared   with the proposed Design 3 of the proposed reversible BCD  adder  as among the proposed design it has minimal number of anicilla inputs , garbage outputs, quantum cost and the delay. }}\\
\hline
\end{tabular}
}
}
\end{table*}

\begin{table*}[!ht]
\centering{
\small{
\caption{ Garbage outputs comparison of n digit reversible BCD adders}
\label{tabSR21DGAR}
\begin{tabular}{|l|l|l|l|p{1.4cm}|p{1.4cm}|}
\hline
Digits & 1 & 2 & \small{Proposed*} &\% Impr. w.r.t 1  & \% Impr. w.r.t 2  \\
\hline
 8 & 32 & 48 & 7  &78.12& 85.4 \\
\hline
 16 &64 & 96 & 15  &76.56&84.37\\
\hline
32 & 128 & 192 & 31& 75.78&83.85 \\ 
\hline
64 & 256 & 384 & 63 &75.39 &83.59  \\ 
\hline
128 & 512 & 768 & 127 &75.19&83.46 \\ 
\hline
256 & 1024 & 1536 & 255 &75.09& 83.39 \\ 
\hline
512 & 2048 & 3072 & 511 & 75.04& 83.36  \\ 
\hline
\multicolumn{6}{|l|}{ \scriptsize{ 1 is the design  in \cite{Thomsen}}}\\
\multicolumn{6}{|l|}{ \scriptsize{ 2 is the design 3 in \cite{majid_bcd2009}}}\\
\multicolumn{6}{|p{8cm}|}{ \scriptsize{ *  is  our  design 3 proposed in this work. In improvement calculation all the existing  designs are compared   with the proposed Design 3 of the proposed reversible BCD  adder  as among the proposed design it has minimal number of anicilla inputs , garbage outputs, quantum cost and the delay. }}\\
\hline
\end{tabular}
}
}
\end{table*}

\begin{table*}[!ht]
\centering{
\small{
\caption{ Quantum cost comparison of n digit reversible BCD adders}
\label{tabSR21DQAC}
\begin{tabular}{|l|l|l|l|p{1.4cm}|p{1.4cm}|}
\hline
Digits & 1 & 2 & \small{Proposed*} &\% Impr. w.r.t 1  & \% Impr. w.r.t 2  \\
\hline
 8 & 1352 & 824 & 560  &58.57& 34.88 \\
\hline
 16 &2704 & 1648 & 1120  &58.57&32.03\\
\hline
32 & 5408 & 3296 & 2240& 58.57&32.03 \\ 
\hline
64 & 10816 & 6592 & 4480 &58.57 &32.03  \\ 
\hline
128 & 21632 & 13184 & 8960 &58.57&32.03 \\ 
\hline
256 & 43264& 26368 & 17920 &58.57& 32.03 \\ 
\hline
512 & 86528 & 52736 & 35840 & 58.57& 32.03  \\ 
\hline
\multicolumn{6}{|l|}{ \scriptsize{ 1 is the design  in \cite{Thomsen}}}\\
\multicolumn{6}{|l|}{ \scriptsize{ 2 is the design 3 in \cite{majid_bcd2009}}}\\
\multicolumn{6}{|p{8cm}|}{ \scriptsize{ *  is  our  design 3 proposed in this work. In improvement calculation all the existing  designs are compared   with the proposed Design 3 of the proposed reversible BCD  adder  as among the proposed design it has minimal number of anicilla inputs , garbage outputs, quantum cost and the delay. }}\\
\hline
\end{tabular}
}
}
\end{table*}

\subsection{ Design of Reversible BCD Adder Based on Approach 2}
In order to design  the reversible  BCD adder based on approach 2, we need a 4 bit reversible adder and a 5 bit reversible binary to BCD converter as illustrated in Fig.\ref{4bitbcdadder1}. 

\subsubsection{ Design 3 of reversible BCD adder with input carry}

\begin{figure*}[!ht]
 \begin{center}
\subfigure [Proposed design 3 of $n$ digit reversible BCD adder with input carry]
   {\label{ndigitd3}
   \includegraphics[width=4.9in]{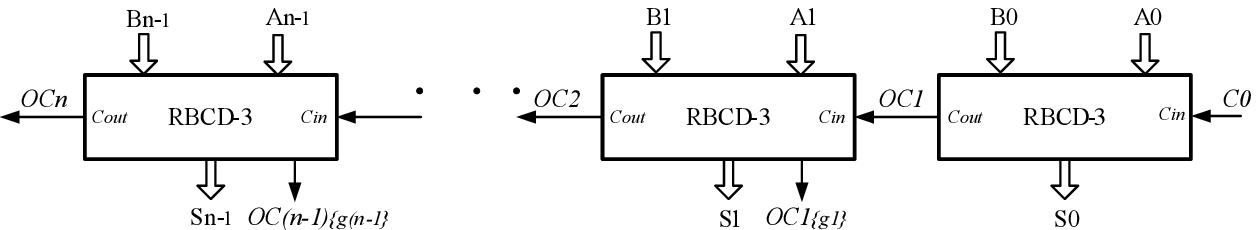}}
   \quad   
\subfigure [Proposed design 4 of $n$ digit reversible BCD adder with no input carry (RBCD-4)]
   {\label{ndigitd4}
   \includegraphics[width=4.9in]{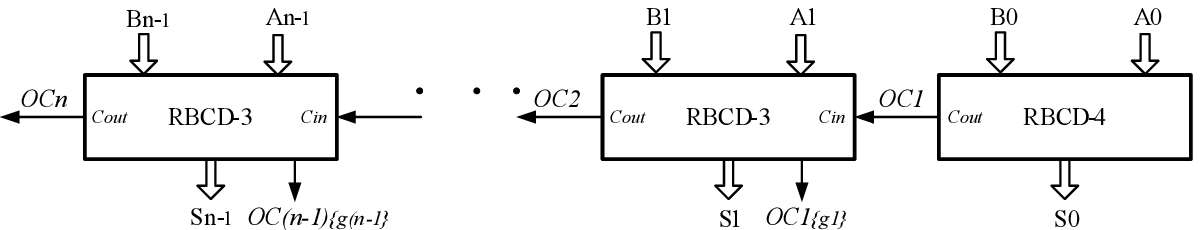}}
   \quad   
 \end{center}
 \caption{ Proposed designs of $n$ digit reversible BCD adder \newline based on approach 2}
 \label{BCDAdder34design}
\end{figure*}

We first present the design of the 1 digit reversible BCD adder with input carry. The 4 bit reversible ripple carry adder with input carry shown in Fig.\ref{4bitadder1ancilla} is used in the design. Recently, an efficient reversible binary to BCD converter without any ancilla bit is proposed in \cite{majid_bcd2009} which we have used in our design. The reversible binary to BCD converter proposed in \cite{majid_bcd2009} is illustrated in Fig.\ref{BintoBCD} and has the quantum cost of 16 and delay of 16 $\Delta$. The proposed design of the 1 digit reversible BCD adder with input carry ($c_0$)   is shown in Fig.\ref{RBCDAdder2inputc} and will be used with name RBCD-3 in this work. The design has 1 ancilla input bit and zero garbage outputs. The quantum cost of the proposed reversible BCD adder with input carry is 70 while the delay is 57 $\Delta$.  \emph{Once we have designed the 1 digit reversible BCD adder (RBCD-3), the $n$ digit reversible BCD adder with input carry can be designed by cascading of the 1 digit reversible BCD adder in the ripple carry fashion as illustrated in Fig.\ref{ndigitd3}}. Thus, the design 3 of the $n$ digit reversible BCD adder with input carry has $n$ ancilla inputs bits, $n-1$ garbage outputs, quantum cost of $70n$ and delay of  $57n \Delta$. As shown in Fig.\ref{ndigitd3} the design 3 of the $n$ digit reversible BCD adder with input carry has $n-1$ garbage outputs as the first 1 digit reversible BCD adder will have no garbage output while the remaining $n-1$ 1 digit reversible BCD adders each will have 1 garbage output. This is because as the output carry of a BCD adder will work as the input carry to the next one and  will be regenerated at the outputs. These regenerated input carries at the outputs will not be used further in the computation and hence will form garbage bits.  Table \ref{tabSR2123}  illustrates that the proposed design is better than the existing design in terms of number of ancilla input bits and garbage outputs while also being efficient in terms of the quantum cost and the delay.

\subsubsection{ Design 4 of reversible BCD adder with no input carry}
\begin{figure}[!ht]
\begin{center}
 \includegraphics[width=2.0in]{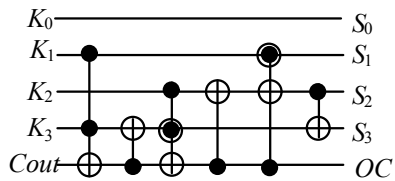}
\end{center}
\caption {Design of binary to BCD Ccnverter [Mohammadi et al. 2009]}
 \label{BintoBCD}
  \end{figure}

\begin{figure}[!ht]
\begin{center}
 \includegraphics[width=4.0in]{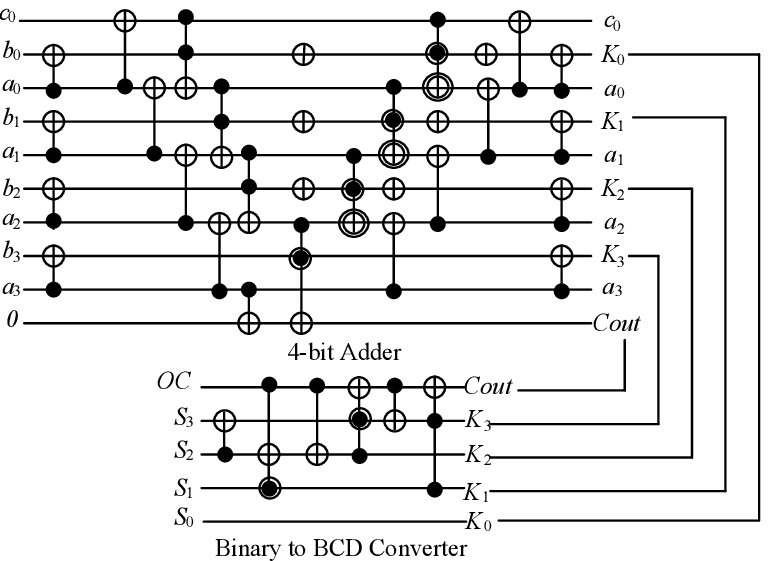}
\end{center}
\caption {Proposed  design of reversible BCD adder with input carry(RBCD-3) based on approach 2. The design consists of 4 bit reversible binary adder with input carry illustrated in Fig.\ref{4bitadder1ancilla} and reversible binary to BCD converter illustrated in Fig. \ref{BintoBCD}. $g_1,g_2,...g_5$ represent the 5 garbage outputs. There is  one ancilla input with constant value of 0}
\label{RBCDAdder2inputc}
  \end{figure}

\begin{figure}[!ht]
\begin{center}
\includegraphics[width=4.0in]{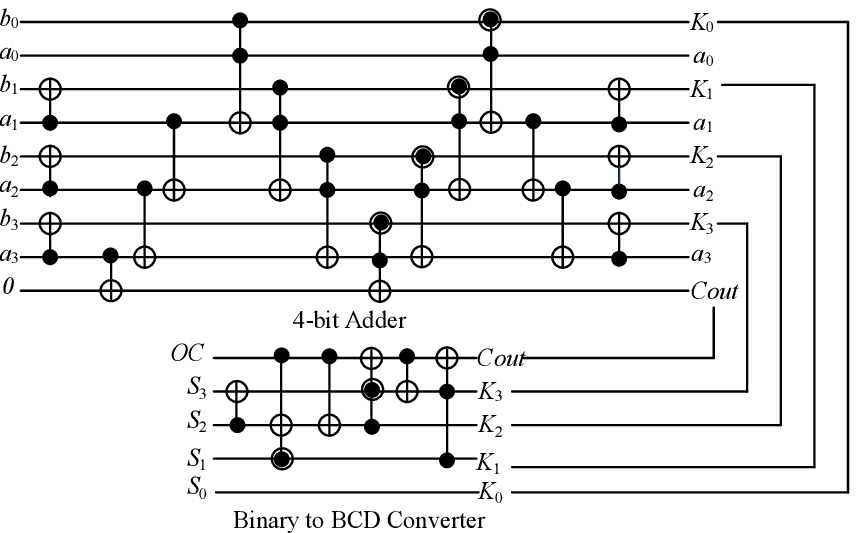}
\end{center}
\caption {Proposed  design of reversible BCD adder with no input carry (RBCD-4) based on approach 2. The design consists of 4 bit reversible binary adder with no input carry illustrated in Fig.\ref{4bitadder0ancilla} and reversible binary to BCD converter illustrated in Fig. \ref{BintoBCD}. $g_1,g_2,...g_4$ represent the 4 garbage outputs. There is  one ancilla input with constant value of 0}
\label{RBCDAdder2noinputc}
  \end{figure}

We are also proposing another design of the n digit reversible BCD adder that has no input carry. We first design the 1 digit reversible BCD adder with no input carry which is shown in Fig.\ref{RBCDAdder2noinputc}, and will be used with name RBCD-4 further in this work. The design uses the 4 bit reversible ripple carry adder with no input carry illustrated in Fig.\ref{4bitadder0ancilla} along with the design of the reversible binary to BCD converter illustrated in Fig.\ref{BintoBCD}. The proposed design of 1 digit reversible BCD adder has 1 ancilla input bit and has zero garbage outputs. The quantum cost of the design is 62 while the propagation delay is 54 $\Delta$. Once we have designed the 1 digit reversible BCD adder with no input carry, the $n$ digit reversible BCD adder with no input carry can be designed by utilizing the 1 digit reversible BCD adder with no input carry (RBCD-4) to add the least significant digit and then cascading  $n-1$  1 digit reversible BCD adder with input carry (RBCD-3) in the ripple carry fashion. The design of n digit reversible BCD adder with no input carry is illustrated in Fig.\ref{ndigitd4}. Thus, the design 4 of $n$ digit reversible BCD adder without input carry has  $n$ ancilla inputs bits, $n-1$ garbage outputs, quantum cost of $70n-8$ and delay of  $57n-3 \Delta$. Table \ref{tabSR2123}  illustrates that the proposed design is better than the existing design in terms of number of ancilla input bits and the garbage outputs while also being efficient in terms of the quantum cost and the delay.

\subsection{Comparison of n digit Reversible BCD Adders}
All the existing designs of the reversible BCD adders are with input carry $c_0$. Among our proposed design of $n$ digit  reversible BCD adder with input carry, the design 3 has the minimum number of ancilla inputs bits, garbage outputs, quantum cost and the delay. The results of all the existing works and our proposed work are summarized in  Table \ref{tabSR2123}.  Among the existing works shown in Table \ref{tabSR2123} for the design of $n$ digit reversible BCD adder,  the design presented in \cite{Thomsen} has the minimum number of garbage outputs, while the design 3 presented in \cite{majid_bcd2009} has the minimum number of ancilla inputs. Thus we have shown the comparison of our proposed work with the design presented in \cite{Thomsen} and \cite{majid_bcd2009} in Tables \ref{tabSR21DANC}, \ref{tabSR21DGAR} and \ref{tabSR21DQAC} in terms of number of ancilla inputs, number of garbage outputs, and the quantum cost, respectively for values of n ranging from n=8 digits to n=512 digits.  It is to observed that delays of  \cite{Thomsen} and  design 3 of \cite{majid_bcd2009} are not known, thus we are not able to compare our design with these designs in terms of delay. From Table \ref{tabSR21DANC}, it can be observed that in terms of number of ancilla inputs, the  proposed design 3 achieves the improvement ratios of 75\% and  50\% compared to the design presented in \cite{Thomsen}  and \cite{majid_bcd2009}, respectively.  The  improvement ratios in terms of number of garbage outputs range from 78.12\% to 75.04\%, and  85.4\% to 83.36\%  compared to the design presented in \cite{Thomsen} and \cite{majid_bcd2009}, respectively, the details are illustrated in Table \ref{tabSR21DGAR}. As illustrated in Table \ref{tabSR21DQAC}, the  improvement ratios in terms of quantum cost are 58.57\%, and range from 34.88\% to 32.03\% compared to the design presented in  \cite{Thomsen} and \cite{majid_bcd2009}, respectively.
Thus the proposed designs of reversible BCD adders are efficient in terms of number of ancilla inputs, garbage outputs, quantum cost and the delay compared to the existing designs in literature.

\section{Simulation and Verification}
\begin{figure}[!ht]
 \begin{center}
   \includegraphics[width=3.5in]{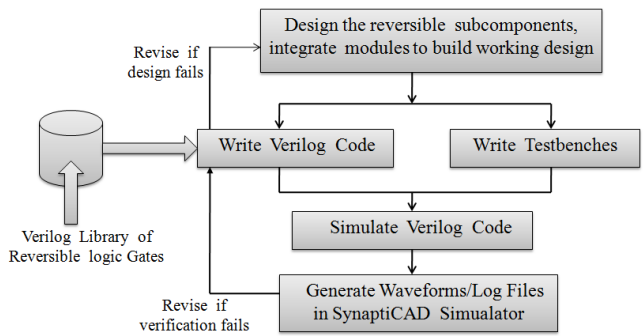}
    \end{center}
 \caption{Simulation flow of reversible circuits using Verilog HDL}
 \label{dfw}
\end{figure}

The proposed reversible ripple carry adder designs, reversible detection unit, reversible correction units, reversible binary to BCD converter and the  complete working designs of the reversible BCD adders are functionally verified through simulations. The simulation is performed by creating a library of reversible gates in Verilog Hardware Description Language and is used  to code the proposed reversible designs. The Verilog library contains the Verilog codes of reversible gates such as the Fredkin gate, the Toffoli gate, the Peres gate, the TR gate, the Feynman gate etc. All the reversible designs of the adders and the subcomponents are coded in Verilog HDL by utilizing the reversible gates from the Verilog library of reversible gates. The test benches are created for every reversible  circuits proposed in this work and for 4 bit and 8 bit reversible binary adders, and 1 digit reversible BCD adders exhaustive simulations are done  to verify the correctness.  The simulation flow used in this work is illustrated in Fig.\ref{dfw}. The ModelSim and the SynaptiCAD simulators are used for the functional verification of the Verilog HDL codes. The waveforms are generated using the SynaptiCAD Verilog simulator.

\section{Conclusions}
In this work, we have presented  efficient designs of  reversible ripple carry binary and BCD adders primarily optimizing the parameters of number of ancilla input bits and the garbage outputs.  The optimization of the quantum cost and the delay are also considered. The reversible designs of subcomponents used in the BCD adder design such as detection unit, correction unit and the binary to BCD converter are also illustrated.  The proposed reversible binary and BCD adders designs are shown to be better than the existing designs in terms of the number of ancilla inputs bits and the garbage outputs while maintaining the lower quantum cost and the delay. We conclude that the use of the specific reversible gates for a particular combinational function can be very much beneficial in minimizing the number of ancilla input bits, garbage outputs, quantum cost and the delay. All the proposed reversible designs are functionally verified at the logical level by using the Verilog hardware description language and the HDL simulators. The proposed efficient designs of  reversible binary and BCD adders will find applications in quantum/reversible computing requiring BCD arithmetic units. 

\subsection*{Acknowledgements}
We would like to express our sincere thanks to the anonymous reviewers for their critical suggestions which helped in improving the manuscript.

\bibliographystyle{acmtrans}
\bibliography{instructions}


\end{document}